\documentclass[journal]{IEEEtran}

\usepackage{amssymb}
\usepackage{amsmath,yhmath}
\usepackage{amsfonts}
\usepackage{amsthm}
\usepackage{empheq}
\usepackage{bigints}
\usepackage[bottom]{footmisc}
\usepackage[inline]{enumitem}
\usepackage[american]{babel}
\usepackage{diagbox}
\usepackage[mathscr]{euscript}
\usepackage{graphicx,scalerel,xcolor}
\usepackage{pgfplots,pgfplotstable,relsize}
\usepackage{tikz,tikzscale}
\usetikzlibrary{arrows.meta}
\usetikzlibrary{patterns}
\usepackage[algo2e]{algorithm2e}

\ifCLASSOPTIONcompsoc
  \usepackage[caption=false,font=normalsize,labelfont=sf,textfont=sf]{subfig}
\else
  \usepackage[caption=false,font=footnotesize]{subfig}
\fi

\usepackage{upgreek}
\usepackage{cite}
\usepackage[normalem]{ulem}
\usepackage{float}

\usepackage{calligra}
\usepackage[T1]{fontenc}

\DeclareMathAlphabet{\mathcalligra}{T1}{calligra}{m}{n}

\pgfplotsset{compat=newest}
\usetikzlibrary{shapes,shadows,arrows.meta,intersections}
\DeclareMathAlphabet\CMath{U}{cmath}{m}{n}

\usetikzlibrary{calc,arrows}
\pgfplotsset{compat=1.16}
\theoremstyle{plain}
\newtheorem{theorem}{Theorem}

\newtheorem{remark}{Remark}

\definecolor{asparagus}{rgb}{0.53, 0.66, 0.42}
\definecolor{cadmiumgreen}{rgb}{0.0, 0.42, 0.24}

\definecolor{revisionorange}{RGB}{210,105,0}

\newcommand{\iu}{\textrm{j}}
\newcommand{\vect}[1]{\boldsymbol{\mathbf{#1}}}

\newcommand{\Ttran}{\mathsf{T}}
\newcommand{\Her}{\mathsf{H}}

\begin{document}

\title{Grid Design for Near-Field Channel Estimation with Uniform Linear Arrays}
\author{Antonio A. D'Amico,~\IEEEmembership{Senior~Member,~IEEE}, and L. Antonelli}
\maketitle

\begin{abstract}
This paper develops deterministic grids for line-of-sight near-field channel
estimation with uniform linear arrays (ULAs) and hybrid receivers. A normalized
channel approximation squared error (NCASE) criterion leads to geometric
covering problems under a constant-amplitude parabolic model and an
elliptical approximation of the correlation regions. For a horizontal ULA,
transformation and scaling yield a disk-covering problem, addressed by
adapting a hexagonal lattice to the mapped ground-plane region. For a
vertical ULA, a reciprocal-distance interval cover is solved optimally by
a farthest-reaching greedy procedure. Numerical evaluation with exact
spherical-wave channels yields empirical maximum approximation errors close
to the target and improved P-SOMP estimation relative to the tested
polar-domain grids of comparable size.
\end{abstract}

\section{Introduction}
The increasing demand for high-data-rate wireless communications has
motivated the development of multiple-input multiple-output (MIMO)
systems operating at millimeter-wave (mmWave) and terahertz (THz)
frequencies~\cite{Castro2020,Huang_MM_2011,Rappaport_2019}. These bands
provide abundant bandwidth but also suffer from severe propagation losses,
making directional transmission and reception through large antenna arrays
particularly important. As array sizes increase, fully digital beamforming,
with one radio-frequency (RF) chain per antenna, may entail excessive
hardware cost, power consumption, and calibration complexity
\cite{Ning_OJCS_2023}.

Hybrid analog--digital architectures provide a compromise between hardware
complexity and beamforming flexibility~\cite{molisch2017hybrid}. A limited
number of RF chains interface with the antenna array through an analog
network, while the baseband processor operates on a reduced-dimensional
signal. Depending on the interconnection topology, each RF chain is connected
either to all antenna elements or to a subarray
\cite{heath2016overview,el2014spatially}.

A central issue in hybrid MIMO systems is the efficient
representation and estimation of the wireless channel. At high
carrier frequencies, the propagation channel is often dominated by a
limited number of significant paths and can therefore be represented
by a small number of steering vectors associated with suitable
propagation parameters. In grid-based approaches, the continuous
parameter space is discretized into a finite set of candidate points,
forming a \emph{grid} and its associated \emph{dictionary}. The
choice of this grid directly affects channel estimation, beam
training, and hybrid precoding. A coarse grid reduces computational
and storage requirements at the price of increased basis mismatch,
whereas an excessively dense grid increases complexity and may also
lead to highly coherent dictionary columns.

Grid and codebook design have consequently received considerable attention.
Under the far-field plane-wave model, a critically sampled discrete Fourier
transform (DFT) grid provides an orthogonal spatial representation and
enables fast Fourier transform (FFT) processing. DFT-like codebooks also
underlie codebook structures employed in 5G New Radio
\cite{huang2020dfthc,3GPP38214,giordani2019nr}. Hierarchical codebooks reduce
beam-search complexity~\cite{xiao2016codebook,qi2022hierarchical}, while
non-uniform angular quantization improves angular coverage
\cite{chen2018nonuniform}. Dictionary-learning and machine-learning methods
adapt the representation to the propagation environment
\cite{liu2021codebook,bhuyan2024adaptive}. Sparse recovery algorithms exploit
the resulting discrete channel representation~\cite{lee2016channel}, and
grid-refinement and off-grid techniques mitigate the mismatch with continuous
propagation parameters~\cite{tang2019offgrid}.

As antenna apertures become electrically large, the far-field
approximation ceases to be accurate over an increasingly large
spatial region. In the resulting near-field regime, the
spherical-wave response depends jointly on direction and distance,
and both phase and amplitude variations across the aperture may
become relevant~\cite{LuZeng_TWC2022}. Near-field focusing also
exhibits a finite depth in the radial dimension, whose dependence on
the array aperture and geometry has been characterized in
\cite{KosasihBjornson_TWC2024}. These properties motivate channel
dictionaries that account explicitly for both angular and radial
variations.

Several near-field dictionary and codebook constructions have been proposed.
Cui and Dai~\cite{CuiDai_TCOM2022} introduced a polar-domain representation
with uniform sampling of the spatial directional cosine and non-uniform
distance sampling based on a dictionary-column coherence criterion. Demir and
Bj\"ornson~\cite{10403506} developed a polar-domain dictionary for extremely
large aperture planar arrays, also using non-uniform distance sampling to
reduce column coherence. Wu and Dai~\cite{10123941} proposed a
spherical-domain codebook for location division multiple access, exploiting
near-field resolution in both angle and distance. In
\cite{zhang2024sparse}, distance is instead an unknown dictionary parameter
within an angular-domain sparse representation. Related work addresses
near- and far-field codebook design and hierarchical beam alignment
\cite{zhang2024codebook,chen2023hierarchical}, as well as fast near-field
beam training and low-complexity hierarchical codebooks
\cite{zhang2022fast,zhu2025clusterbased}.

A particularly relevant geometric approach is due to Zheng et al.
\cite{Zheng_TCOM2025}. For a uniform linear array (ULA), their construction
uses a correlation threshold, congruent elliptical approximations in a
transformed parameter domain, and staggered sampling associated with
hexagonal cells. These features are closely related to the horizontal ULA
(HULA) construction considered here. The present HULA development focuses
on target-specific calibration of the ellipse axes using the finite-array
correlation, explicit mapping of the ground-plane user region, and selection
of lattice translation and orientation over its bounded image.
Section~II gives the coordinate and threshold correspondence and clarifies
the scope of these differences.

In previous work, we investigated near-field dictionary design for
physically constrained user geometries. An error-based polar-domain design
for elevated ULAs was developed in~\cite{11443816}, while
\cite{antonelli2026novelnearfielddictionarydesign} considered planar arrays
with users confined to, or close to, a reference plane. The present work
differs from~\cite{11443816}, which tunes angular and distance sampling for
a prescribed dictionary size, by determining the number and placement of
atoms from a target approximation accuracy and by treating vertical ULAs
(VULAs) as well.

Motivated by these considerations, this paper develops a deterministic
grid-design framework that links a prescribed channel-approximation
accuracy to the geometry of the array and the physical region of interest.
Unlike the coherence-driven sampling in~\cite{CuiDai_TCOM2022}, the design
is formulated as a geometric covering problem. This distinction also
applies at zero array height. For a HULA, a constant-amplitude parabolic
channel model and an elliptical approximation of the correlation regions
lead, after transformation and scaling, to a disk-covering problem.
A hexagonal lattice is adapted to the finite transformed region through
a search over translation and orientation, yielding a low-cardinality
lattice-based grid.

For a VULA, cylindrical symmetry leads to a one-dimensional covering
problem in reciprocal distance. Under a sufficient condition, the
admissible set associated with each grid point is an interval. An exchange
argument then establishes that a farthest-reaching greedy procedure gives
a minimum-cardinality solution of the resulting interval-covering problem.
A simplified construction with uniform spacing in reciprocal distance is
also considered. A separate numerical comparison gives virtually identical
maximum and average approximation errors for the two constructions with
the considered parameters; the reported VULA error curves use the greedy
construction.

The geometric design uses tractable channel and correlation approximations,
whereas numerical validation uses exact spherical-wave channels with
element-dependent amplitudes. The evaluation examines the approximation
errors and the effect of the HULA dictionaries on single-path channel
estimation using polar-domain simultaneous orthogonal matching pursuit
(P-SOMP). This connects the grid-design criterion to its use in a hybrid
receiver, where dictionary size affects storage and sparse-recovery
complexity.

The remainder of this paper is organized as follows.
Section~II develops the grid design for a horizontal ULA and maps the
approximation requirement into a covering problem in a transformed
domain. Section~III considers a vertical ULA and derives the
corresponding one-dimensional interval-covering formulation.
Section~IV describes the hybrid receiver signal model and the
P-SOMP channel-estimation procedure used for performance evaluation.
Section~V presents the numerical results, and Section~VI concludes
the paper.

\section{Grid Design for a Horizontal Uniform Linear Array}
\label{Sec:GridDesignHULA}
\begin{figure}[t]
\centering
\begin{tikzpicture}[scale=0.75, transform shape]
    \def\xscale{0.707106781186548} 
    
     \draw [line width=1.5 pt]  (-2,-0.1) -- (-2,0.1) node [anchor=south]{\scriptsize $1$};
     \draw [line width=1.5 pt]  (-1.5,-0.1) -- (-1.5,0.1);
     \draw [line width=1.5 pt]  (-1,-0.1) -- (-1,0.1);
     \draw [line width=1.5 pt]  (1,-0.1) -- (1,0.1);
     \draw [line width=1.5 pt]  (1.5,-0.1) -- (1.5,0.1);
     \draw [line width=1.5 pt]  (2,-0.1) -- (2,0.1) node [anchor=south]{\scriptsize $M$};
     
     \draw [{Stealth}-{Stealth}] (-1.5, -0.5) -- (-1,-0.5);
     \node at (-1.25,-0.75) {\scriptsize $\delta$};
     \draw [line width=0.5 pt, dashed,] (-1.5, 0) -- (-1.5,-1);
     \draw [line width=0.5 pt, dashed,] (-1.0, 0) -- (-1.0,-1);
     
     \draw [line width=1 pt, dashed]  (-0.8,0) -- (0.8,0);
     \draw [line width=0.01 pt]  (-2,0) -- (2,0);

    \draw [thin] [{}-{Stealth}] (0,0) -- (0-2.5*\xscale,0-2.5*\xscale) node  [anchor=north east]{\scriptsize $X$}; 
    \draw [thin] [{}-{Stealth}] (0,0) -- (3,0) node [anchor=west]{\scriptsize $Y$};
    \draw [thin][{}-{Stealth}]  (0,0) -- (0,3) node [anchor=south]{\scriptsize $Z$};
    
    \node[] at (1/6 ,1/4 ) {\scriptsize $\mathbf O$};
    
    \node[] at (1/4 ,-2.8+1/4 ) {\scriptsize $\mathbf O'$};
    
    \node[right] at (-6.5,-5.5) { \textsf{Ground Plane}};
    \draw [] (-7, -6) rectangle (1,-5.8);
    \draw [line width=0.5 pt]  (-4,-2.8) -- (4,-2.8);
    \draw [line width=0.5 pt]  (4,-2.8) -- (4,-3);
    \draw [line width=0.5 pt]  (-4,-2.8) -- (-7,-5.8);
    \draw [line width=0.5 pt]  (4,-2.8) -- (1,-5.8);
    \draw [line width=0.5 pt]  (4,-3) -- (1,-6);
    \draw [{Stealth}-{Stealth}] (0, 0) -- (0,-2.8);
    \node[right] at (0,-1.4) {\scriptsize $b$};
    
    \draw[fill = gray] plot[variable=\t, domain=-180:0, samples=100] ({sqrt(2)*sin(\t) - 2*cos(\t)}, {sqrt(2)*sin(\t) - 2.8});
    \draw (0,-2.8) -- (-1.4142,-1.4142-2.8);
    \node[right] at (-0.7,-3.5) {\scriptsize $\rho_{\min}$};
       
\end{tikzpicture}
\caption{Geometry of the system with a horizontal array.}
\label{fig:Geometry}
\end{figure}
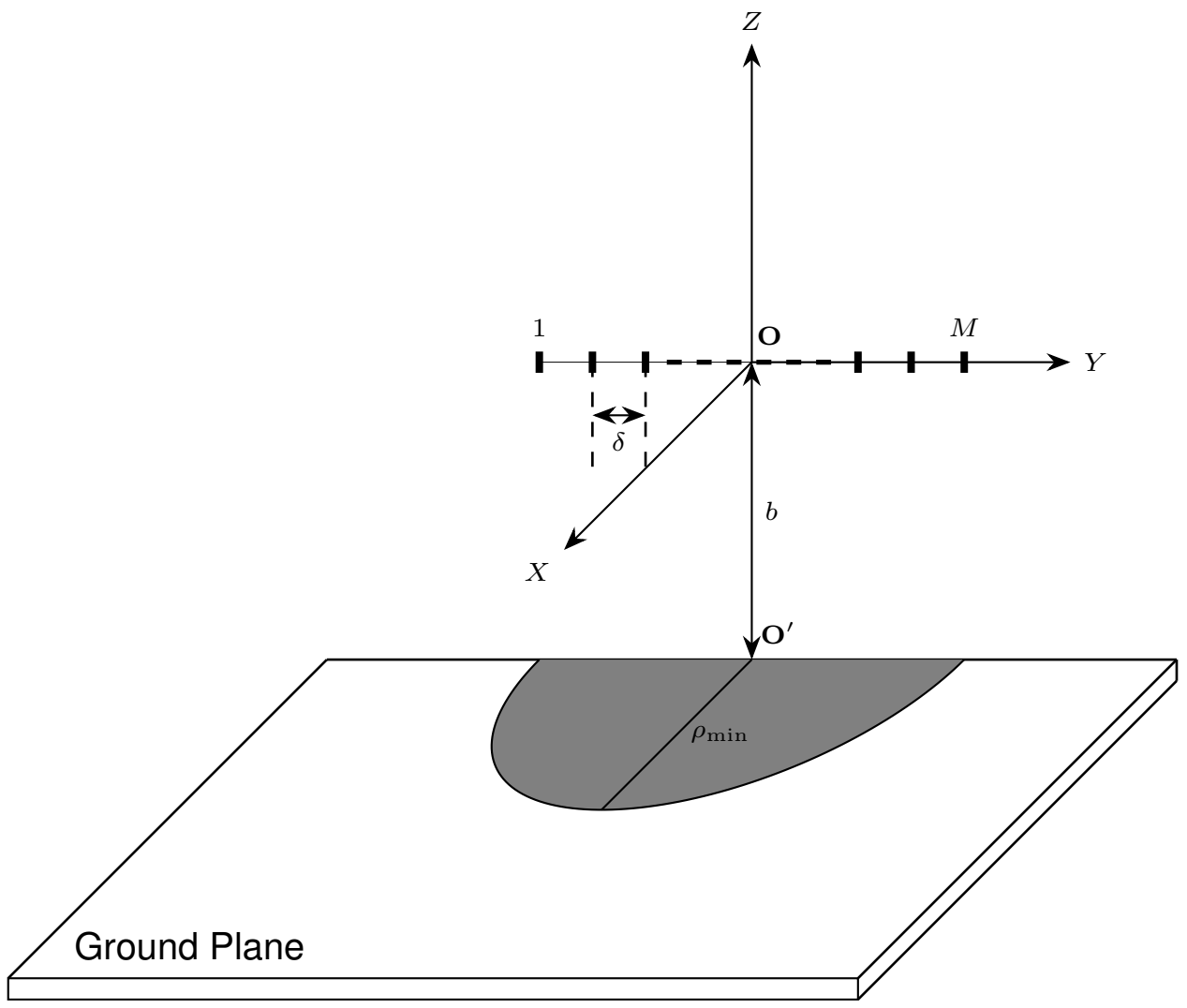

We consider a base station (BS) equipped with a ULA of $M$ antennas with inter-element spacing $\delta$, positioned at height $b$ above the \textit{ground plane} (GP), as shown in Fig.~\ref{fig:Geometry}. For notational convenience, we assume $M$ is odd; the derivations can be readily extended to even $M$. We first consider the case of a horizontal array in which the antennas are arranged along the $y$-axis of a Cartesian coordinate system, whose origin $\vect{O}$ coincides with the center of the array. The ground plane is assumed parallel to the $xy$-plane. The analysis will then be extended to a vertical array whose elements are arranged along the $z$-axis.

With the above assumptions, the coordinates of the $m$th antenna, $\mathbf{u}_m = (x_m, y_m, z_m)$, $m = 1,\dots,M$, are
\begin{equation}
\label{eq:AntCoor}
x_m = z_m = 0, \quad y_m = i(m)\delta, \quad i(m) = m - 1 - \frac{M - 1}{2}\raisebox{2 pt}{.}
\end{equation}
A point on the GP is denoted by $\mathbf{p} = (p_x, p_y, p_z)$ with
\begin{equation}
p_x = \rho_\mathrm{p} \cos \varphi_\mathrm{p}, \quad p_y = \rho_\mathrm{p} \sin \varphi_\mathrm{p}, \quad p_z = -b
\end{equation}
where $(\rho_\mathrm{p},\varphi_\mathrm{p})$ are its polar coordinates. The UEs are assumed to
lie in a physical region on the GP, which defines the \textit{region-of-interest} (RoI). 
For analytical convenience, we construct the grid for the conservative region $\mathcal{R} = \left\{(\rho,\varphi) | \varphi \in [-\pi/2,\pi/2],\rho \ge \rho_{\min}\right\}$, where $\rho_{\min}$ is chosen so that the physical RoI is contained in \(\mathcal R\). Therefore, any grid satisfying the prescribed approximation criterion over \(\mathcal R\) also satisfies it over the original physical RoI,
possibly at the price of using more grid points than strictly necessary for
the finite physical sector.
This is illustrated in Fig.~\ref{fig:Geometry} in which the filled area represents the \textit{forbidden region} of the ground plane. 

Under line-of-sight (LoS) propagation conditions, the channel between $\mathbf{p}$ and the $m$th antenna is
\begin{equation}
h_{m}(\mathbf{p}) = \sqrt{\xi_m} \, e^{-j\frac{2\pi}{\lambda} r_m}
\label{eq:channel_model}
\end{equation}
where $\xi_m$ accounts for path loss and antenna gains, $r_m = \|\mathbf{p} - \mathbf{u}_m\|$ is the Euclidean distance, and $\lambda$ is the wavelength. The channel vector
\begin{equation}
\mathbf{h}(\mathbf{p}) = \left[ h_{1}(\mathbf{p}), \dots, h_{M}(\mathbf{p}) \right]^\mathsf{T} \in \mathbb{C}^M
\end{equation}
can be expressed as
\begin{equation}
\mathbf{h}(\mathbf{p}) = \boldsymbol{\chi}(\mathbf{p}) \odot \mathbf{a}(\mathbf{p})
\label{eq:channel}
\end{equation}
where $\boldsymbol{\chi}(\mathbf{p}) = \left[\sqrt{\xi_1}, \dots, \sqrt{\xi_M}\right]^\mathsf{T}$ is the \emph{amplitude vector}, $\mathbf{a}(\mathbf{p}) = \left[ e^{-j\frac{2\pi}{\lambda}r_1}, \dots, e^{-j\frac{2\pi}{\lambda}r_M} \right]^\mathsf{T}$ is the \emph{array steering vector}, and $\odot$ denotes the Hadamard product.

Let $\mathcal G$ be a finite set of physical grid points. We seek the
smallest $|\mathcal G|$ such that, for every $\mathbf p\in\mathcal R$,

\begin{equation}
\label{eq:DesignCriterion}
\operatorname{NCASE}(\mathbf{p}) \le \epsilon
\end{equation}
where
\begin{equation}
\label{eq:NMSE_Mod}
\operatorname{NCASE}(\mathbf{p}) \overset{\textrm{def}}{=} \underset{\mathbf{g} \in \mathcal{G}}{\min} \ \underset{\alpha}{\min} \ \frac{\| \mathbf{h}(\mathbf{p}) - \alpha \mathbf{h}(\mathbf{g}) \|^2}{\| \mathbf{h}(\mathbf{p}) \|^2}
\end{equation}
denotes the normalized channel approximation squared error (NCASE),
where $\alpha\in\mathbb C$ and $\epsilon\in(0,1)$ is a prescribed target.
For each user position, the complex scale factor removes differences in
overall gain and phase, so that the criterion measures the residual error
in the spatial channel response. Minimization over $\mathcal G$ selects
the dictionary atom that gives the smallest relative squared error.
The grid-design problem is therefore

\begin{equation}
\label{eq:DesignCriterionGeneral1}
\begin{array}{rl}
      & \underset{\mathcal{G}}{\min} \, |\mathcal{G}|    \\
      \mathrm{s.t.} &  \operatorname{MNCASE} \le \epsilon
\end{array}
\end{equation}
where
\begin{equation}
\label{eq:NCASE_General}
\operatorname{MNCASE} = {\underset{\mathbf{p} \in \mathcal{R}}{\sup}} \operatorname{NCASE}(\mathbf{p})\raisebox{2 pt}{.}
\end{equation}

The supremum accounts for the unbounded design region $\mathcal R$;
the numerical results use maxima over finitely many sampled positions.
The corresponding average over the user distribution is

\begin{equation}
\label{eq:ANCASE}
\operatorname{ANCASE} = \mathbb{E}_{\mathbf{p}} \left\{ \operatorname{NCASE}(\mathbf{p}) \right\}\raisebox{2pt}{.}
\end{equation}
The minimization with respect to $\alpha$ in~\eqref{eq:NMSE_Mod} is straightforward and yields
\begin{equation}
\label{eq:alfa_opt}
\widehat{\alpha} = \frac{\mathbf{h}^{{\mathsf H}}(\mathbf{g}) \, \mathbf{h}(\mathbf{p})}{\| \mathbf{h}(\mathbf{g}) \|^2}\raisebox{2pt}{.}
\end{equation}
Substituting~\eqref{eq:alfa_opt} into~\eqref{eq:NMSE_Mod} gives
\begin{equation}
\label{eq:NMSE_Opt}
\operatorname{NCASE}(\mathbf{p}) = 1 - \underset{\mathbf{g} \in \mathcal{G}}{\max} \ \frac{\left| \mathbf{h}^{\mathsf H}(\mathbf{g}) \, \mathbf{h}(\mathbf{p}) \right|^2}{\| \mathbf{h}(\mathbf{p}) \|^2 \, \| \mathbf{h}(\mathbf{g}) \|^2}
\end{equation}
so that \eqref{eq:DesignCriterion} can be equivalently written as
\begin{equation}
\label{eq:DesignCriterion2}
\underset{\mathbf{g} \in \mathcal{G}}{\max} \ \frac{\left| \mathbf{h}^{\mathsf H}(\mathbf{g}) \, \mathbf{h}(\mathbf{p}) \right|^2}{\| \mathbf{h}(\mathbf{p}) \|^2 \, \| \mathbf{h}(\mathbf{g}) \|^2} \ge 1 - \epsilon.
\end{equation}

Approximating the amplitudes as constant across the array,
$\boldsymbol\chi(\mathbf p)\approx\sqrt{\xi_p}\mathbf1$ and
$\boldsymbol\chi(\mathbf g)\approx\sqrt{\xi_g}\mathbf1$, with $\mathbf1$
the all-ones vector, gives

\begin{equation}
\label{eq:DesignCriterion3}
\frac{\left| \mathbf{h}^{\mathsf H}(\mathbf{g}) \, \mathbf{h}(\mathbf{p}) \right|^2}{\| \mathbf{h}(\mathbf{p}) \|^2 \, \| \mathbf{h}(\mathbf{g}) \|^2} \approx \dfrac{\left| \mathbf{a}^{\mathsf H}(\mathbf{g}) \, \mathbf{a}(\mathbf{p}) \right|^2}{M^2}\raisebox{2 pt}{.}
\end{equation}

We use this constant-amplitude approximation for grid design.

Next, we adopt the second-order (parabolic) approximation of the distance $r_m = \| \mathbf{p} - \mathbf{u}_m \|$:
\begin{equation}
\label{eq:ParabolicAppx}
\begin{split}
r_m &\approx R - (\delta \cos \theta \sin\varphi) i(m) + \dfrac{\delta^2}{2R} \left( 1 - \cos^2 \theta  \sin^2 \varphi \right) i^2(m) 
\end{split}
\end{equation}
\noindent where $R = \sqrt{\rho^2 + b^2}$ is the distance from $\mathbf{p}$ to the center of the array, and $\theta$ is the elevation angle, with $\cos \theta = \sqrt{1 - b^2R^{-2}}$.
Based on \eqref{eq:ParabolicAppx}, after straightforward computations, the squared modulus of the correlation on the right-hand side of \eqref{eq:DesignCriterion3} can be approximated as \cite[eq. (9)]{CuiDai_TCOM2022}:
\begin{equation}
\label{eq: Correlation1}
\dfrac{ \left| \mathbf{a}^{\mathsf H}(\mathbf{g}) \, \mathbf{a}(\mathbf{p}) \right|^2}{M^2} \approx \left | \dfrac{1}{M}\sum_{m = - (M-1)/2}^{(M-1)/2} e^{\iu (A m + B m^2)}\right |^2 \overset{\textrm{def}}{=} \Lambda(A,B)
\end{equation}
where
\begin{equation}
\label{eq: A}
A = \dfrac{2 \pi \delta} {\lambda} (\Gamma_p -\Gamma_g)
\end{equation}
\begin{equation}
\label{eq: B}
B = \dfrac{\pi \delta^2} {\lambda} \left( \eta_g - \eta_p \right)
\end{equation}
with 
\begin{equation}
\label{eq: Gamma}
\Gamma_i \overset{\textrm{def}}{=} \cos \theta_i \sin \varphi_i=\sqrt{1 - b^2R_i^{-2}}\sin \varphi_i    \quad \quad i = p,g   
\end{equation}
and
\begin{equation}
\label{eq: eta}
\eta_i \overset{\textrm{def}}{=} (1 - \Gamma_i^2) R_i^{-1} \quad \quad i = p,g  \raisebox{2 pt}{.}
\end{equation}
\begin{remark}[Relation with existing near-field designs]
Identifying the directional cosine and range in~\cite{CuiDai_TCOM2022}
with $\Gamma$ and $R$, their rings
$r_{s,n}=Z_\Delta(1-\theta_n^2)/s$ become the separable sampling
$\Gamma_n=\theta_n$, $\eta_s=s/Z_\Delta$, with $\eta=0$ for the far-field ring.
For Zheng's half-wavelength ULA~\cite{Zheng_TCOM2025},
$\beta=\Gamma$ and $\alpha=\lambda\eta/4$. With
$\delta_\alpha=\alpha_p-\alpha_g$ and
$\delta_\beta=\beta_p-\beta_g$, we have
$A=\pi\delta_\beta$, $B=-\pi\delta_\alpha$, and
$\Lambda(A,B)=f^2(\delta_\alpha,\delta_\beta)$, where $f$ is their
correlation magnitude. Thus $c=\sqrt{1-\epsilon}$ under the common
constant-amplitude parabolic model. The correlation and NCASE targets,
as well as the elliptical and staggered geometry, are therefore closely
related. The present HULA construction uses target-specific finite-array
ellipse intercepts, the mapped ground-plane design region, and a finite
search over lattice translation and orientation.
\end{remark}

\begin{figure}[!t]
\centering
\subfloat[$10\log_{10}\epsilon=-3$ dB.]{%
\includegraphics[width=0.8\columnwidth]{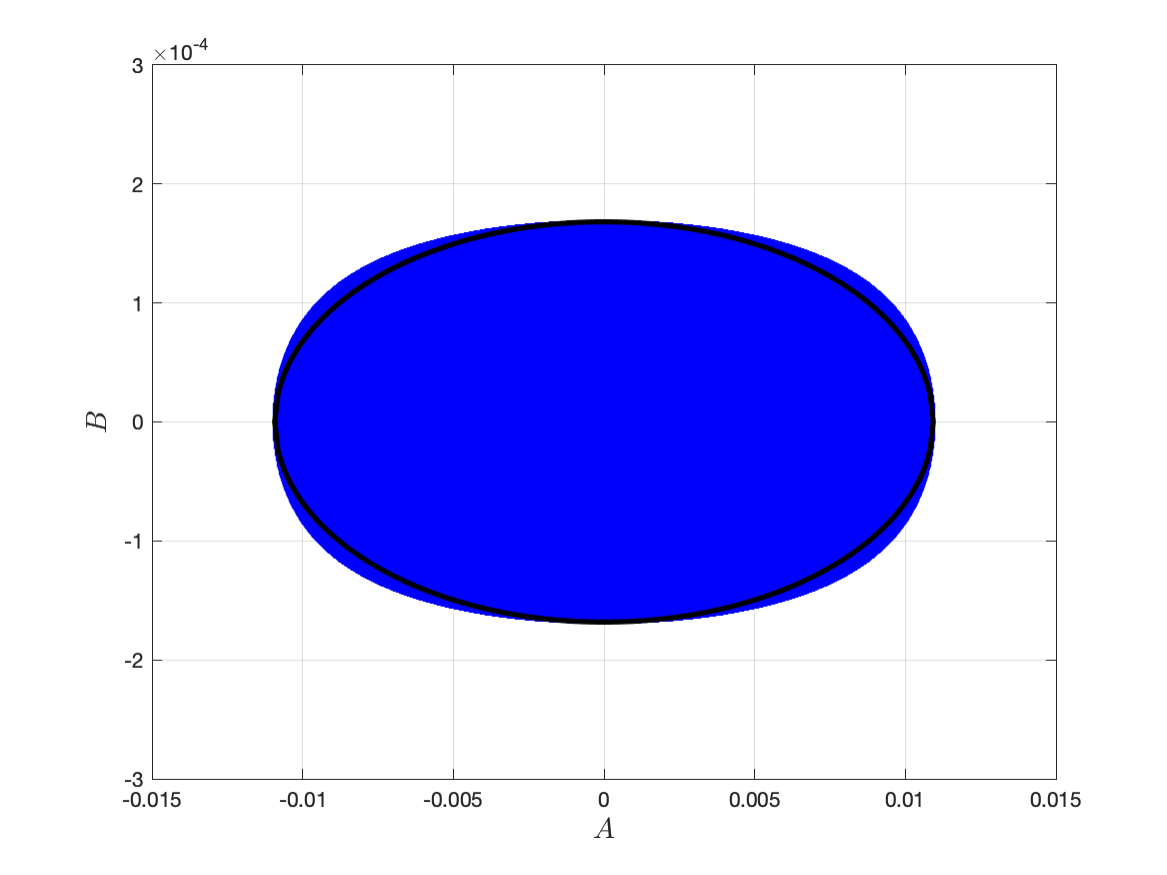}
\label{fig:AR_t_3}
}

\par
\subfloat[$10\log_{10}\epsilon=-5$ dB.]{%
\includegraphics[width=0.8\columnwidth]{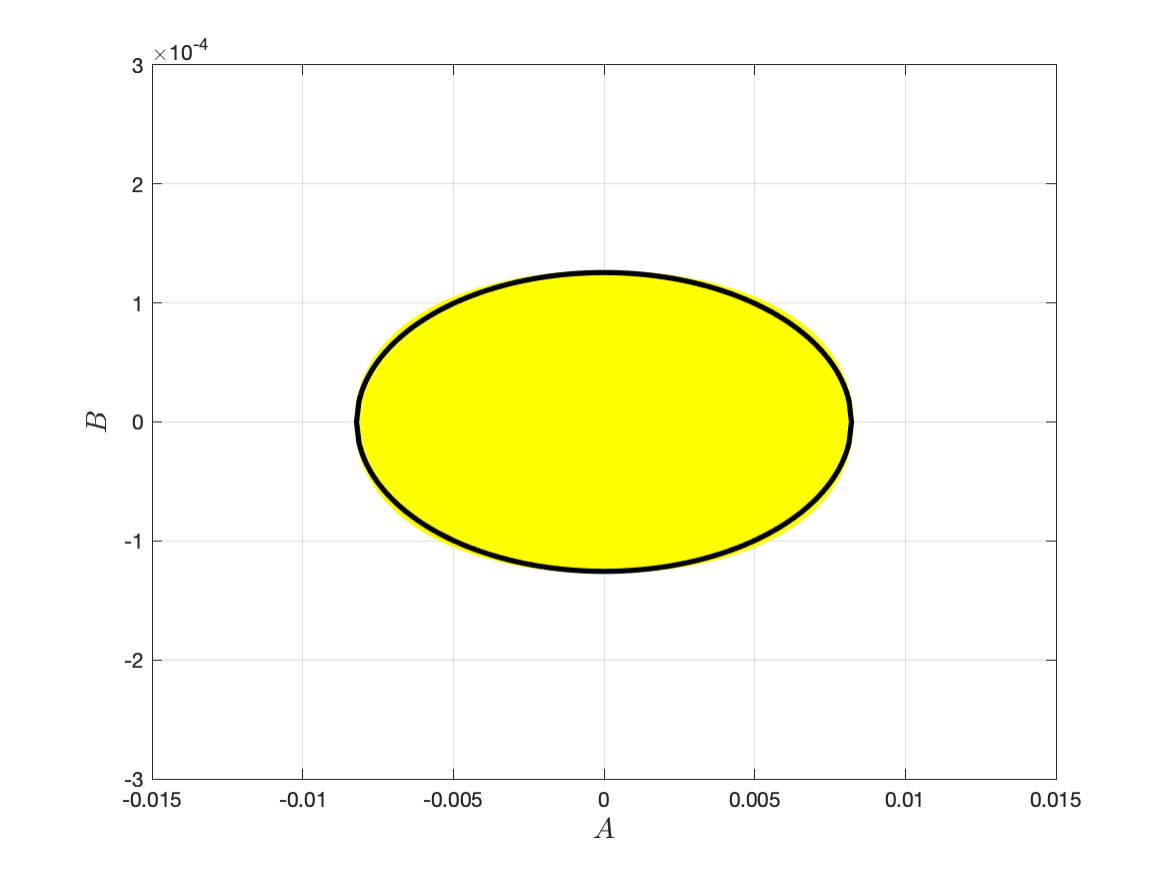}
\label{fig:AR_t_5}
}
\caption{Admissible region in the $(A,B)$-domain for different values of $\epsilon$.}
\label{fig:AdmissibleRegions}
\end{figure}

Figs.~\ref{fig:AdmissibleRegions}\subref{fig:AR_t_3}
and~\ref{fig:AdmissibleRegions}\subref{fig:AR_t_5} show the regions of the $(A,B)$-plane in which 
\begin{equation}
\label{eq:LevelRegion}
\Lambda(A,B) \ge 1 - \epsilon
\end{equation}
for $10\log_{10}\epsilon=-3$ and $-5$~dB. Since $\Lambda$ is
$2\pi$-periodic in both arguments, we use only the central component
containing the origin and do not exploit nonlocal replicas.

Appendix~\ref{App:AdmissRegion} motivates the elliptical approximation

\begin{equation}
\label{eq:Ellipse}
\dfrac{A^2}{\widehat{A}^2_{\epsilon}} + \dfrac{B^2}{\widehat{B}^2_{\epsilon}} = 1
\end{equation}
where $\widehat{A}_{\epsilon}$ and $\widehat{B}_{\epsilon}$, respectively, are the smallest positive solutions to $\Lambda(A,0) = 1 - \epsilon$ and $\Lambda(0,B) = 1 - \epsilon$.

\begin{table}[t!]
  \centering
\begin{tabular}{|c|c|c|c|}
\hline \\ [-6 pt]
$10 \log_{10}\epsilon$ & $\widehat{A}_{\epsilon}$ & $\widehat{B}_{\epsilon}$ & $\widehat{A}_{\epsilon}/\widehat{B}_{\epsilon}$\\ [3 pt]
\hline
     -20 & +1.3605e-03 & +2.0672e-05 &    65.81 \\ 
     -15 & +2.4300e-03 & +3.6941e-05 &    65.78 \\ 
     -10 & +4.3860e-03 & +6.6725e-05 &    65.73 \\ 
      -5 & +8.2155e-03 & +1.2556e-04 &    65.43 \\  
      -3 & +1.0931e-02 & +1.6821e-04 &    64.98 \\ \hline 
\end{tabular}
  \caption{Values of $\widehat{A}_{\epsilon}$, $\widehat{B}_{\epsilon}$, and of the ratio $\widehat{A}_{\epsilon}/\widehat{B}_{\epsilon}$, for five different values of $10 \log_{10}\epsilon$ and $M = 255$.}
    \label{tab:table0}
\end{table}

As illustrated in Figs.~\ref{fig:AdmissibleRegions}\subref{fig:AR_t_3}
and~\ref{fig:AdmissibleRegions}\subref{fig:AR_t_5}, the calibrated ellipse
closely follows the central correlation contour for the plotted targets.
Table~\ref{tab:table0} gives the axis intercepts for $M=255$ and five
values of $\epsilon$. Their ratio varies only slightly and remains close
to $65$, although the ellipse size changes with the target error.
We use the closed region bounded by~\eqref{eq:Ellipse} as an
admissible-region surrogate. This is an approximation to the correlation
region, rather than an exact error bound for the spherical-wave channel.

Using~\eqref{eq: A}--\eqref{eq: B}, the surrogate maps into congruent
closed ellipses in the $(\Gamma,\eta)$-plane:

\begin{equation}
\label{eq:EllipseGEta}
\mathsf E_g(\epsilon):
\frac{(\Gamma-\Gamma_g)^2}{\widehat C_\epsilon^2}
+
\frac{(\eta-\eta_g)^2}{\widehat D_\epsilon^2}
\le 1
\end{equation}
where 
\begin{equation}
\label{eq:GammaHat}
\widehat{C}_{\epsilon} = \dfrac{\lambda}{2 \pi \delta} \widehat{A}_{\epsilon} 
\end{equation}
and
\begin{equation}
\label{eq:etaHat}
\widehat{D}_{\epsilon} = \dfrac{\lambda}{\pi \delta^2} \widehat{B}_{\epsilon}\raisebox{2 pt}{.}
\end{equation}

Under the adopted approximations, an ellipse centered at
$(\Gamma_g,\eta_g)$ describes the parameter pairs represented by its atom
within the target error. Covering the transformed user region with these
ellipses therefore converts the approximation requirement into a geometric
constraint on the grid.

Let $\mathcal R'$ be the closure of the image of $\mathcal R$ under
\eqref{eq: Gamma}--\eqref{eq: eta}. Taking the closure includes the limiting
parameter values as the physical distance tends to infinity. Let
$\mathcal G'\subset\mathbb R^2$ be a finite set of ellipse centers.
Unlike the physical grid $\mathcal G$, these centers need not lie in
$\mathcal R'$ or have a ground-plane counterpart: each still defines
a dictionary atom through~\eqref{eq:SteeringVectorApp}. The resulting
surrogate covering problem is

\begin{equation}
\label{eq:DesignCriterionHULA_2}
\begin{aligned}
\underset{\mathcal G'}{\min}\quad
& |\mathcal G'| \\
\mathrm{s.t.}\quad
& \bigcup_{(\Gamma_g,\eta_g)\in\mathcal G'}
  \mathsf E_g(\epsilon)\supseteq\mathcal R'.
\end{aligned}
\end{equation}

We approximate this finite covering problem by restricting the centers
to translated and rotated hexagonal lattices.

The boundary of $\mathcal R'$ follows directly from the physical
constraints. For $b>0$ and $|\Gamma|<1$, the requirement
$|\sin\varphi|\le1$ in~\eqref{eq: Gamma}, together with
$\rho\ge\rho_{\min}$, gives
\[
R\ge\max\left\{R_{\min},\frac{b}{\sqrt{1-\Gamma^2}}\right\}
\]
where $R_{\min}=\sqrt{\rho_{\min}^2+b^2}$. At fixed $\Gamma$,
$\eta=(1-\Gamma^2)/R$ decreases with $R$. Its upper boundary is
therefore obtained at the smallest feasible range, while $\eta=0$
is included by closure as $R\to\infty$. The two range constraints
coincide at $|\Gamma|=\rho_{\min}/R_{\min}$, yielding

$$\mathcal{R}' = \big\{(\Gamma,\eta): \Gamma \in [-1,1], \eta \in \left[0, \eta_{\mathrm{u}}(\Gamma)\right]\big\}$$ 
where
\begin{equation}
\label{eq: etau}
\eta_{\mathrm{u}}(\Gamma) = \left\{\begin{array}{cl}  R_{\min}^{-1} (1 - \Gamma^2)  & \text{\hspace{6 pt}for $0 \le \big|\Gamma\big| < \dfrac{\rho_{\min}}{R_{\min}}$ } \\ \vspace{-12pt} \\b^{-1}(1 - \Gamma^2)^{3/2} & \text{\hspace{6 pt}for $\dfrac{\rho_{\min}}{R_{\min}} \le \big|\Gamma \big| \le 1 $ }\end{array}\right.{.}
\end{equation}

\begin{figure}[!t]
\centering
\includegraphics[width=0.8\columnwidth]{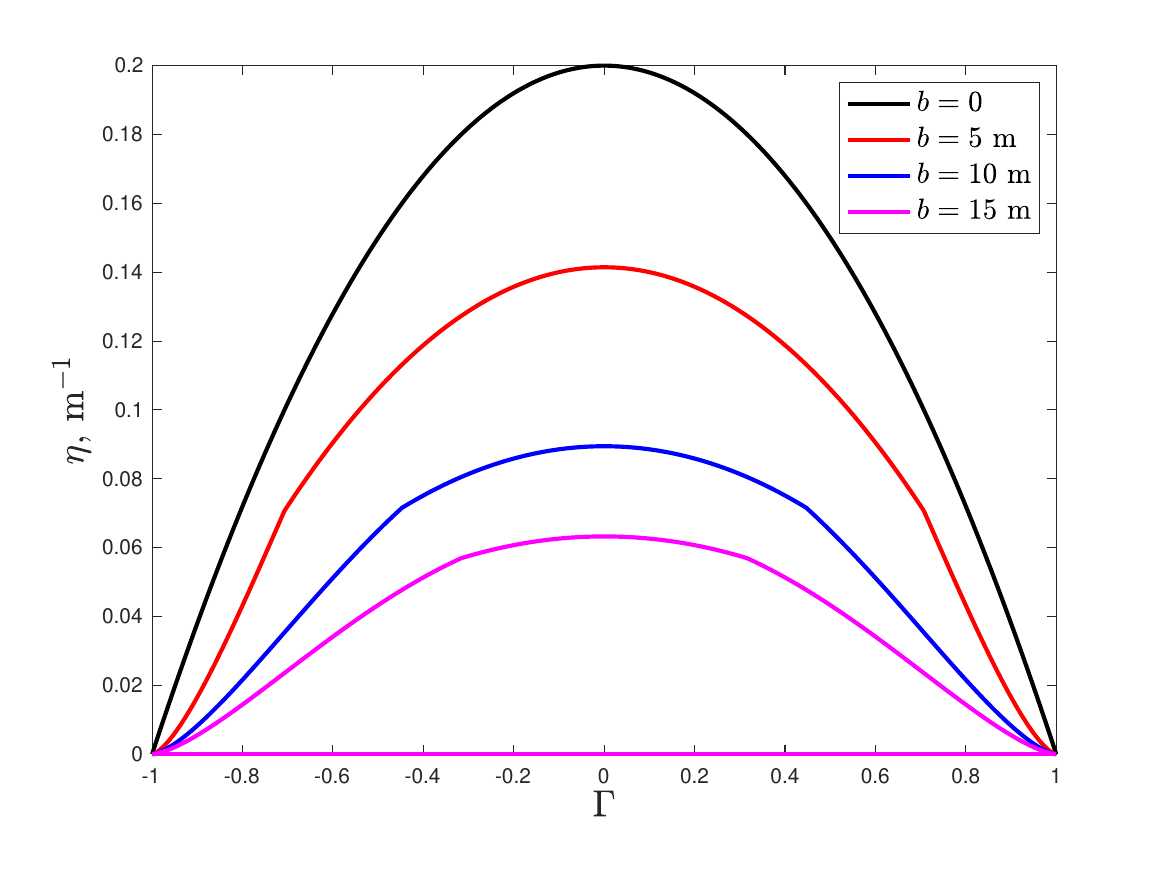}
\caption{Contour of the region-of-interest $\mathcal{R}'$ in the $(\Gamma,\eta)$-domain, for four different values of the array height: $b=0,5,10$, and $15$ m.}
\label{fig: RoI_GammaEta}
\end{figure}

The endpoint values at $\Gamma=\pm1$ are zero. For $b=0$, only
the constraint $R\ge R_{\min}$ remains, so
$\eta_{\mathrm u}(\Gamma)=R_{\min}^{-1}(1-\Gamma^2)$ throughout
$[-1,1]$. Figure~\ref{fig: RoI_GammaEta} illustrates how the mapped
domain contracts as the array height increases.

To construct a practical cover, we scale the transformed coordinates as
\[
\tilde\Gamma=\frac{\Gamma}{\widehat C_\epsilon}\raisebox{2pt}{,}
\qquad
\tilde\eta=\frac{\eta}{\widehat D_\epsilon}\raisebox{2pt}{.}
\]

This one-to-one scaling transforms the congruent ellipses into unit
disks and the region $\mathcal R'$ into
\[
\mathcal R''=\left\{(\tilde\Gamma,\tilde\eta):
|\tilde\Gamma|\le\frac{1}{\widehat C_\epsilon},\quad
0\le\tilde\eta\le
\frac{\eta_u(\widehat C_\epsilon\tilde\Gamma)}{\widehat D_\epsilon}
\right\}.
\]
The scaled center set $\mathcal G''$ satisfies
$|\mathcal G''|=|\mathcal G'|$, so minimizing the number of ellipses is
equivalent to minimizing the number of covering disks.

Appendix~\ref{App:Solution} constructs a practical cover by restricting
the disk centers to the hexagonal lattice in \eqref{eq:HexagonalLattice}. The translation and orientation
affect how the lattice meets the boundary of the finite region and hence
the number of selected centers. We retain the placement giving the
smallest count over a finite search set. Thus, optimality is confined to
the tested placements, and the construction is not asserted to solve the
unrestricted finite-cardinality covering problem. Since the dictionary is
designed offline, this search does not increase the online channel-estimation
complexity.

We test $N'=60$ deterministic uniformly spaced orientations $\omega\in[0,\pi/3)$ and, for each orientation, $N''=100$
translations represented by deterministic points uniformly distributed in the reference
hexagon $\mathcal H_0^\omega$. For each triplet
$(\tilde\Gamma_0,\tilde\eta_0,\omega)$, we construct the corresponding cover
of $\mathcal R''$ and retain the one with the smallest number of selected disks.

Table~\ref{tab:table2} shows the values of \((\tilde\Gamma_0,\tilde\eta_0,\omega)\) that give the smallest number of grid points among the considered lattice translations and rotations, for four different values of the array height $b$. For the considered parameters, \(N_{\rm G}\) decreases as
the array height increases. This behavior is consistent with
the progressive contraction of the mapped region
\(\mathcal R'\) in the \(\eta\) direction as \(b\) increases,
as also illustrated in Fig.~\ref{fig: RoI_GammaEta}.
The other system parameters are reported in Table~\ref{tab:SystemParameters}.
\begin{table}[t!]
  \centering
    \begin{tabular}{c|c|c|c|c}
      Array height $b$, m & $\tilde\Gamma_0^{\rm opt}$ & $\tilde\eta_0^{\rm opt}$ & $\omega_{\rm opt}, \operatorname{rad}$ & $N_\mathrm{G}$\\
      \hline
      0 & $\sqrt{3}/2$ & 1/2& $\pi/6$ & 1659 \\
      5 & $\sqrt{3}/2$ & 1/2& $\pi/6$ & 1224 \\
      10 & $\sqrt{3}/2$ & 1/2& $\pi/6$  & 860 \\
      15 & $\sqrt{3}/2$ & 1/2 & $\pi/6$  & 695 \\
    \end{tabular}
  \caption{Best values of \((\tilde\Gamma_0,\tilde\eta_0,\omega)\) over the considered
lattice translations and rotations, and corresponding number \(N_\mathrm{G}\)
of grid points, for different values of \(b\).}
    \label{tab:table2}
\end{table}

\begin{table}[h!]
\caption{Simulation Parameters}
\label{tab:SystemParameters}\centering 
\begin{tabular}{|c|c|}
\hline
\textbf{Parameter} & \textbf{Value} \\ \hline\hline
Carrier Frequency, $f_c$ & $30\,\, [\mathrm{GHz}]$ \\ \hline
Array Aperture, $L_H$ & $1.27$ [m]\\ \hline
Number of Antennas, $M$ & $255$ \\ \hline
Antenna Spacing, $\delta$ & $\lambda/2$ \\ \hline
$\rho_{\min}$ & $5$ [m] \\ \hline
$\rho_{\max}$ & $50$ [m] \\ \hline
$\varphi_{\min}$ & $-\pi/3$ [rad] \\ \hline
$\varphi_{\max}$ & $\pi/3$ [rad] \\ \hline
\end{tabular}
\end{table}

For the selected scaled centers, $k=1,\ldots,N_{\rm G}$, inversion gives

\begin{equation}
\label{eq:HULAGridCenters}
\Gamma_{\mathrm g,k}
=\widehat C_\epsilon\tilde\Gamma_{\mathrm g,k}
\qquad
\eta_{\mathrm g,k}
=\widehat D_\epsilon\tilde\eta_{\mathrm g,k}.
\end{equation}
The resulting parameter grid is
$\mathcal G'=\{(\Gamma_{\mathrm g,k},\eta_{\mathrm g,k}):
k=1,\ldots,N_{\mathrm G}\}$.

The points of $\mathcal G'$ generate the corresponding
\textit{dictionary atoms} (see Section IV-B), namely the steering vectors
\begin{equation}
\label{eq:SteeringVectorApp}
[\mathbf{a}(\Gamma_{\mathrm{g},k},\eta_{\mathrm{g},k})]_m = \exp\left\{\iu\dfrac{2 \pi \delta i(m)}{\lambda} \left[\Gamma_{\mathrm{g},k} - \dfrac{\delta i(m)}{2}\eta_{\mathrm{g},k}\right]\right\}
\end{equation}
defined by the parabolic channel model. A center in the image of the
physical RoI can be associated with a physical UE position; otherwise,
it still defines a valid dictionary atom. For a parameter grid
$\mathcal G'$, the NCASE criterion in \eqref{eq:NMSE_Mod} is interpreted
as minimization over the associated dictionary atoms in place of the
physical channel vectors $\mathbf h(\mathbf g)$.

For a center in the physical image with $\eta_g>0$,
$|\Gamma_g|<1$, and $\rho_g>0$, the inverse map is
\begin{equation}
\label{eq:HULAInversePhysical}
\begin{aligned}
R_g&=\frac{1-\Gamma_g^2}{\eta_g}\\
\rho_g&=\sqrt{R_g^2-b^2}\\
\varphi_g&=\arcsin\!\left(\frac{\Gamma_g R_g}{\rho_g}\right).
\end{aligned}
\end{equation}
The boundary derived above ensures that $\rho_g\ge\rho_{\min}$
and that the arcsine argument lies in $[-1,1]$. This inverse map
need not exist for lattice centers outside the physical image.
Such centers are used directly as phase parameters in
\eqref{eq:SteeringVectorApp}. Likewise, centers with $\eta_g=0$
define plane-wave atoms through the same equation without requiring
a finite range. This distinction allows the cover to include boundary
disks while keeping the dictionary well defined.

\begin{figure}[!t]
\centering
\subfloat[$b=0$.]{%
\includegraphics[width=0.8\columnwidth]{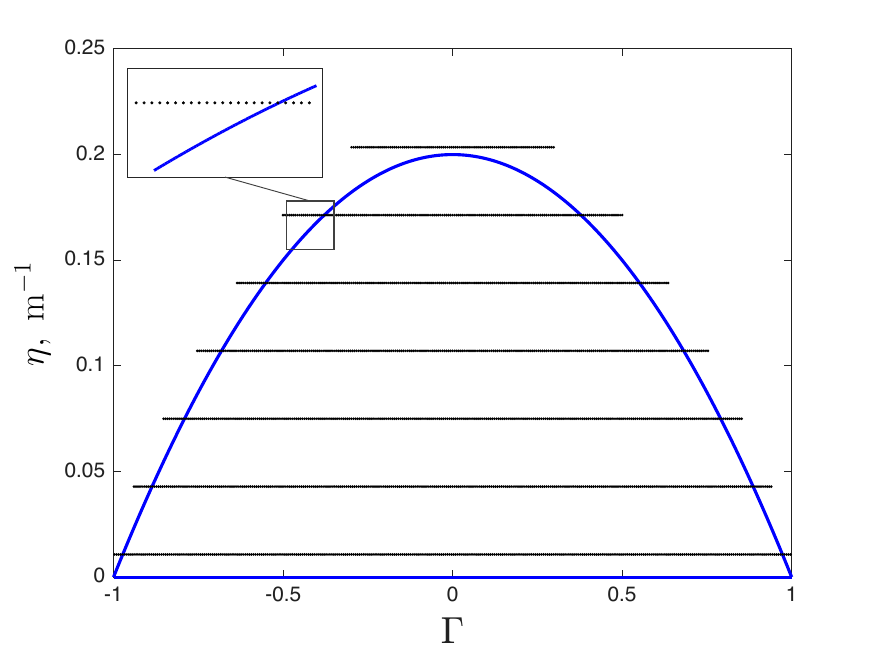}
\label{fig:GP0}}
\par
\subfloat[$b=10$ m.]{%
\includegraphics[width=0.8\columnwidth]{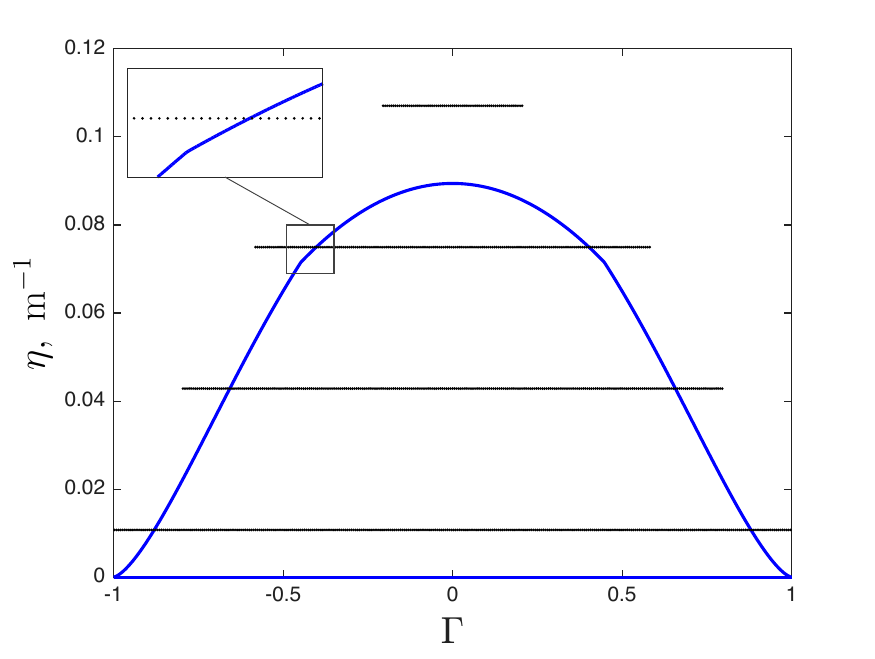}
\label{fig:GP10}}
\caption{Grid points in the $(\Gamma,\eta)$-plane for two values of the array height.}
\label{fig:GridPointsRoI}
\end{figure}

Figs.~\ref{fig:GridPointsRoI}\subref{fig:GP0}
and~\ref{fig:GridPointsRoI}\subref{fig:GP10} show the proposed grids in the \((\Gamma,\eta)\)-plane for \(b=0\) and \(b=10\) m, respectively. The grid points are arranged along straight lines parallel to \(\eta=0\). Because of their small horizontal spacing, the individual markers are difficult to distinguish at the scale of the main plots, giving the appearance of continuous horizontal lines. The enlarged views resolve the individual markers and clarify that each apparent line is a sequence of discrete grid points. For a fixed array configuration and approximation target, the ellipse semi-axes—and hence the lattice spacing—are independent of \(b\), whereas the transformed region \(\mathcal{R}'\) contracts in the \(\eta\) direction as the height increases, as also illustrated in Fig. 3. Consequently, fewer lattice rows are required, and the dictionary cardinality decreases from 1659 atoms at \(b=0\) to 860 atoms at \(b=10\) m, a reduction of approximately \(48.2\%\). This reduction results from the smaller parameter region rather than a relaxation of the design target. The insets also show that some selected centers lie outside \(\mathcal{R}'\), because their associated covering regions intersect its interior. Such centers define valid dictionary atoms through (29), even when they have no corresponding ground-plane position.

To validate the proposed design, we generated $10^6$ UE positions uniformly
with respect to area over the physical~RoI. Although the grid construction is based on 
the constant-amplitude, parabolic, and elliptical approximations
introduced above, the numerical evaluation is performed using the channel
model in \eqref{eq:channel_model}, with exact antenna-to-UE distances and
element-dependent amplitudes. The ANCASE is computed by
averaging the NCASE values in the linear domain over the simulated UE
positions and converting the resulting average to dB.

\begin{figure}[!t]
\centering
\includegraphics[width=0.7 \columnwidth]{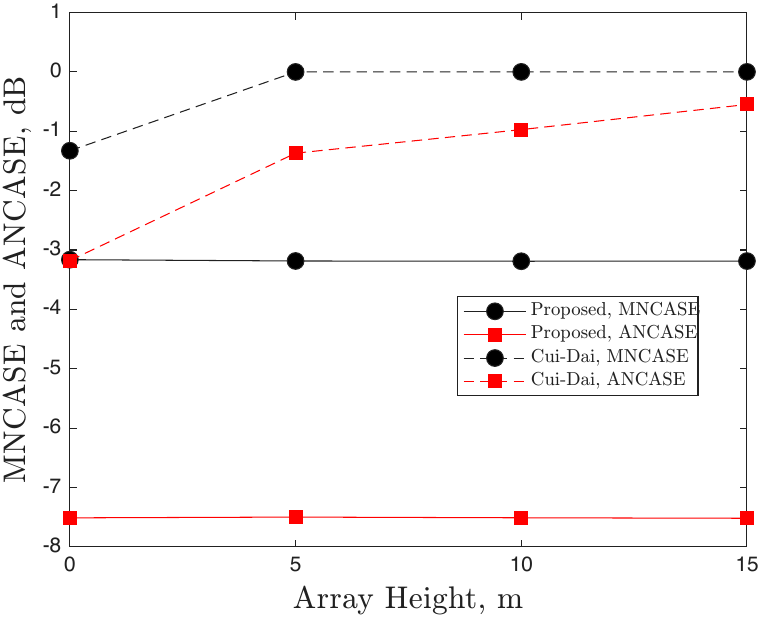}
\caption{MNCASE (black) and ANCASE (red) versus array height for $10^6$ UE positions uniformly distributed with respect to area over the RoI. Dashed curves: Cui--Dai design.}
\label{fig:NCASE_HULA}
\end{figure}

Fig.~\ref{fig:NCASE_HULA} reports
the empirical MNCASE and ANCASE. The empirical MNCASE is essentially
independent of $b$ and remains slightly below the target
$10\log_{10}\epsilon=-3$ dB, whereas ANCASE lies between $-7$ and $-8$ dB.
For comparison, the figure also reports the Cui--Dai grids~\cite{CuiDai_TCOM2022}
with approximately the same numbers of points (1785, 1275, 1020, and 765),
whose approximation errors are larger, particularly for $b\ne0$.

\section{Grid Design for a Vertical Uniform Linear Array}
We now consider a uniform linear array with vertical orientation. In this case, the coordinates of the $m$th antenna, $\mathbf{u}_m = (x_m, y_m, z_m)$, $m = 1,\dots,M$, are
\begin{equation}
x_m = y_m = 0, \quad z_m = i(m)\delta, \quad i(m) = m - 1 - \frac{M - 1}{2}
\end{equation}
and the second-order (parabolic) approximation of the distance $r_m = \| \mathbf{p} - \mathbf{u}_m \|$ reads:

\begin{equation}
\label{eq:ParabolicAppxVULA}
\begin{split}
r_m &\approx  R + \dfrac{\delta b}{R} i(m) + \dfrac{\delta^2}{2R} \left( 1 - \dfrac{b^2}{R^2}\right) i^2(m).
\end{split}
\end{equation}
Based on \eqref{eq:ParabolicAppxVULA}, we can write:
\begin{equation}
\label{eq: Correlation1_VULA}
 \dfrac{\left| \mathbf{a}^{\mathsf H}(\mathbf{g}) \, \mathbf{a}(\mathbf{p}) \right|^2}{M^2} \approx \left | \dfrac{1}{M}\sum_{m = - (M-1)/2}^{(M-1)/2} e^{\iu (A m + B m^2)}\right |^2
\end{equation}
where
\begin{equation}
\label{eq: A_VULA}
A = \dfrac{2 \pi \delta b} {\lambda} (R^{-1}_p -R^{-1}_g)
\end{equation}
\begin{equation}
\label{eq: B_VULA}
B = \dfrac{\pi \delta^2} {\lambda} \left( \xi_p - \xi_g \right)
\end{equation}
with
\begin{equation}
\label{eq: csi}
\xi_i \overset{\textrm{def}}{=} (1 - b^2R_i^{-2}) R_i^{-1} \quad \quad i = p,g.  
\end{equation}
By cylindrical symmetry, the VULA steering response is independent of the
ground-plane azimuth: positions having the same radial distance produce
the same antenna-to-user distances. Consequently, $A$ and $B$ depend only
on $R_p$ and $R_g$. We may select all grid points on the ray
$\varphi_g=0$ without losing coverage of other azimuths in the RoI.

Set $t=R^{-1}$ and $t_g=R_g^{-1}$. The relation
$\xi=t(1-b^2t^2)$ constrains the transformed physical points to a curve in
the $(t,\xi)$-plane. Each admissible set is obtained by intersecting this
curve with the ellipse associated with a grid center and identifying the
corresponding values of $t$. Thus, although the correlation is expressed
in two variables, the grid design has only one independent parameter.
From~\eqref{eq: A_VULA}--\eqref{eq: B_VULA}, the ellipse
\eqref{eq:Ellipse} maps to
\begin{equation}
\label{eq:AdmReg_t_csi}
\left(\frac{t-t_g}{\widehat C_\epsilon}\right)^2
+
\left(\frac{\xi-\xi_g}{\widehat D_\epsilon}\right)^2 \le 1
\end{equation}
where
\begin{equation}
\label{eq:hatChatD}
\widehat{C}_{\epsilon} = \dfrac{\lambda}{2 \pi \delta b} \widehat{A}_{\epsilon} \quad \textrm{and}\quad \widehat{D}_{\epsilon} = \dfrac{\lambda}{\pi \delta^2} \widehat{B}_{\epsilon}.
\end{equation}
Here $b>0$; Appendix~\ref{App:SolutionVULA} treats $b=0$ separately.
In this section, $\widehat C_\epsilon$ is a reciprocal-distance
semi-axis, whereas its HULA counterpart in~\eqref{eq:GammaHat}
is a directional-cosine semi-axis.

Since $t$ and $\xi$ are related through \eqref{eq: csi}, the elliptical region \eqref{eq:AdmReg_t_csi} can be mapped into the admissible set in the \(t\)-domain given by
\begin{equation}
\label{eq:Itg}
\mathcal I(t_g)
=
\left\{
t\in[0,1/b]:\Phi(t;t_g)\le 1
\right\}
\end{equation}
where
\begin{equation}
\label{eq:Ellipse_t_csi}
\Phi(t;t_g) = \left(\frac{t-t_g}{\widehat C_\epsilon}\right)^2
+
\left(\frac{\xi(t)-\xi_g}{\widehat D_\epsilon}\right)^2
\end{equation}
with
\begin{equation}
\label{eq:csi_t}
\xi(t)=t(1-b^2t^2), \quad \xi_g = t_g(1-b^2t^2_g).
\end{equation}
For the VULA, let $\mathcal G'\subseteq\mathcal R'$ denote the finite
set of selected reciprocal distances, where
$\mathcal R'=[t_{\min},t_{\max}]$,
$t_{\min}=R_{\max}^{-1}$, and $t_{\max}=R_{\min}^{-1}$.
The grid-design problem in the reciprocal-distance domain is
\begin{equation}
\label{eq:DesignCriterionVULA_2}
\begin{aligned}
\underset{\mathcal G'\subseteq\mathcal R'}{\min}\quad
& |\mathcal G'| \\
\mathrm{s.t.}\quad
& \bigcup_{t_g\in\mathcal G'}\mathcal I(t_g)
  \supseteq\mathcal R'.
\end{aligned}
\end{equation}
On the selected ray $\varphi_g=0$, each $t_g\in\mathcal G'$ determines
the physical point
$\mathbf g(t_g)=(\sqrt{t_g^{-2}-b^2},0,-b)$.
Thus, $\mathcal G=\{\mathbf g(t_g):t_g\in\mathcal G'\}$ and
$|\mathcal G|=|\mathcal G'|$.

Under the sufficient condition in Appendix~\ref{App:SolutionVULA},
$\mathcal I(t_g)=[t_{\rm start}(t_g),t_{\rm stop}(t_g)]$.
The resulting interval cover admits the following greedy solution.

\begin{theorem}
\label{th:TheoremVULA}
Assume that, for every $t_g\in\mathcal R'$, the admissible set is an
interval $\mathcal I(t_g)=[t_{\rm start}(t_g),t_{\rm stop}(t_g)]$
containing a nonzero neighborhood of $t_g$ relative to $\mathcal R'$,
and that both endpoints are continuous functions of $t_g$.
Set $\tilde t_0=t_{\min}$. At step $k=1,2,\ldots$, let
$\tilde t_{k-1}$ be the covered frontier and define
\begin{equation}
\label{eq:AdmissibleCentersk}
\mathcal A_{k-1}
=\{t_g\in\mathcal R':\tilde t_{k-1}\in\mathcal I(t_g)\}.
\end{equation}
Among all intervals containing that frontier, select one reaching
farthest to the right:
\begin{equation}
\label{eq:RadialSamplingk}
\hat t_{g_k}\in\operatorname*{arg\,max}_{t_g\in\mathcal A_{k-1}}
t_{\rm stop}(t_g).
\end{equation}
Update the frontier as
\begin{equation}
\label{eq:CoveredFrontier}
\tilde t_k=t_{\rm stop}(\hat t_{g_k})
\end{equation}
and stop at the first $K$ with $\tilde t_K\ge t_{\max}$.
Then $\widehat{\mathcal G}'=\{\hat t_{g_1},\ldots,\hat t_{g_K}\}$
is a minimum-cardinality solution of~\eqref{eq:DesignCriterionVULA_2}.
If several maximizers exist, any one may be selected.
\end{theorem}
\begin{proof}
Choose, within each admissible interval, a relatively open neighborhood
of its center. These neighborhoods cover the compact interval
$\mathcal R'$, so a finite subcover exists. Hence
a minimum-cardinality cover exists. At each frontier below $t_{\max}$,
the interval centered at that frontier is feasible and extends to its
right. Moreover, continuity of the endpoints makes $\mathcal A_{k-1}$
a closed subset of $\mathcal R'$. The maximization is therefore attained,
and each step advances the frontier.

Consider a minimum-cardinality cover of the part of $\mathcal R'$
remaining to the right of $\tilde t_{k-1}$. Since this cover is finite
and its intervals are closed, at least one interval
$\mathcal I(t_g^\star)$ contains the frontier and extends to its right.
The greedy interval also contains the frontier and satisfies
\[
t_{\rm stop}(t_g^\star)
\le t_{\rm stop}(\hat t_{g_k}).
\]
Replacing $\mathcal I(t_g^\star)$ with
$\mathcal I(\hat t_{g_k})$ cannot remove coverage to the right of the
frontier: the portion covered there by the former interval is contained
in the latter. All other intervals can remain unchanged. Thus, there
is an optimal completion whose first interval is the greedy choice.
The remaining intervals cover the part beyond $\tilde t_k$, so the
minimum number of intervals still needed decreases by at least one.
Repeating this exchange proves finite termination and shows that the
greedy procedure uses no more intervals than an optimal cover.
Its feasibility follows from
\[
t_{\rm start}(\hat t_{g_k})
\le\tilde t_{k-1}
\le t_{\rm stop}(\hat t_{g_k})=\tilde t_k
\]
which leaves no gap between consecutive intervals. The resulting cover
therefore has minimum cardinality.
\end{proof}

\begin{remark}
For $b>0$, under the condition in Appendix~\ref{App:SolutionVULA},
both $t_{\rm start}(t_g)$ and $t_{\rm stop}(t_g)$ are nondecreasing.
The feasible centers in~\eqref{eq:RadialSamplingk} consequently form
an interval, and its largest element is a maximizer. If this center
is interior to the center domain, it satisfies
\[
t_{\rm start}(t_g)=\tilde t_{k-1}.
\]
Using the dimensionless variables of Appendix~\ref{App:SolutionVULA},
this condition becomes
\[
F(b\tilde t_{k-1};\zeta)=0
\]
where $\zeta=bt_g>b\tilde t_{k-1}$. The required center is obtained
by solving this scalar sixth-order polynomial equation and selecting
the admissible root. If the upper endpoint of the center domain is
feasible, it can be selected directly. For $b=0$, the closed-form
endpoints in Appendix~\ref{App:SolutionVULA} give
\[
\hat t_{g_k}=\min\{\tilde t_{k-1}+\widehat D_\epsilon,t_{\max}\}
\qquad
\tilde t_k=\hat t_{g_k}+\widehat D_\epsilon.
\]
When $t_{\max}>t_{\min}$, the resulting cardinality is
\[
K=\left\lceil\frac{t_{\max}-t_{\min}}{2\widehat D_\epsilon}\right\rceil.
\]
Indeed, each interval covers a length of at most
$2\widehat D_\epsilon$, which gives this lower bound on $K$, and
the greedy construction attains it. Its centers are uniformly spaced
except for a possible adjustment of the last center to $t_{\max}$.
Thus, the center constraint $\widehat{\mathcal G}'\subseteq\mathcal R'$
is respected also at the final step.
\end{remark}

\subsection{Example of grid construction}

We use the parameters in Table~\ref{tab:SystemParameters} and
$10\log_{10}\epsilon=-3$~dB. Table~\ref{tab:table0} gives
$\widehat A_\epsilon=1.093\times10^{-2}$ and
$\widehat B_\epsilon=1.682\times10^{-4}$; hence
$\widehat D_\epsilon=2.142\times10^{-2}$ and
$\widehat C_\epsilon$ follows from~\eqref{eq:hatChatD}.

\begin{figure}[!t]
\centering
\subfloat[$b=0$.]{%
\includegraphics[width=0.8\columnwidth]{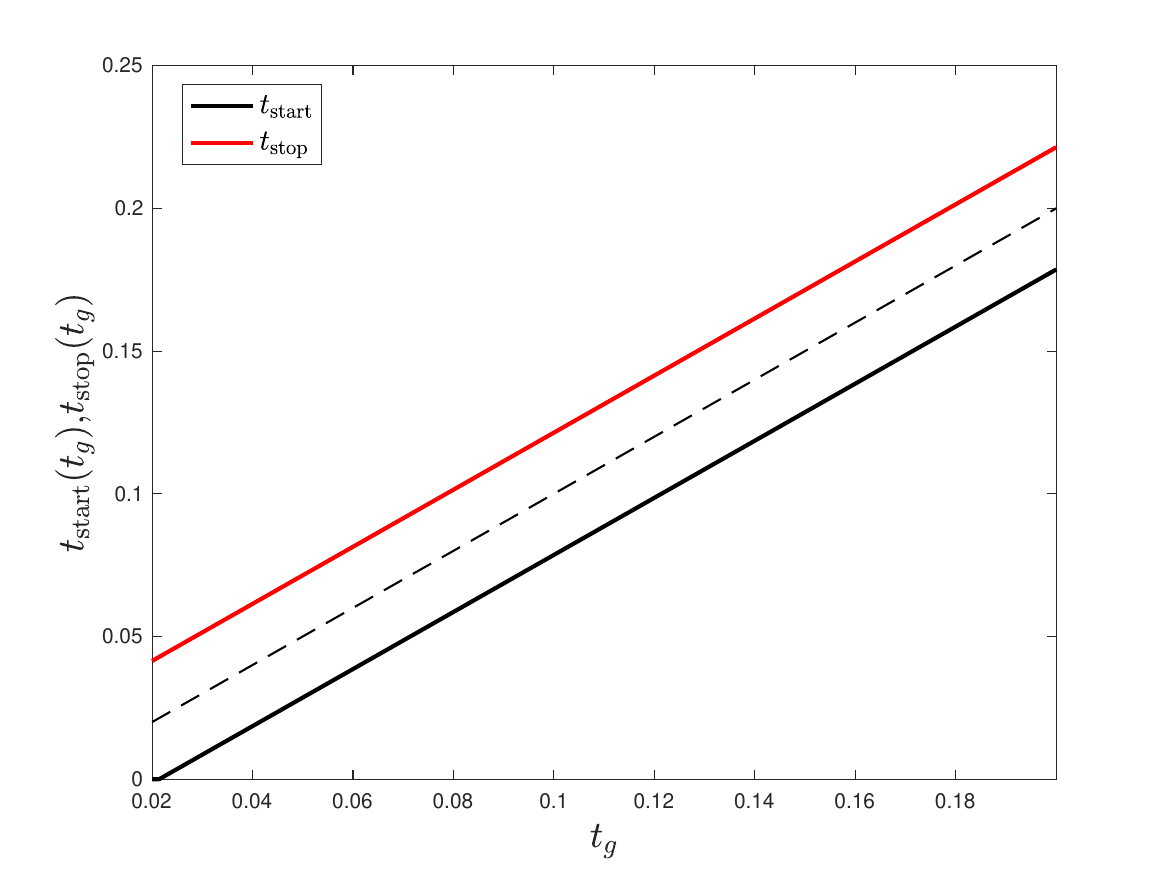}
\label{fig:tstart_tstop_b0}}
\par
\subfloat[$b=10$ m.]{%
\includegraphics[width=0.8\columnwidth]{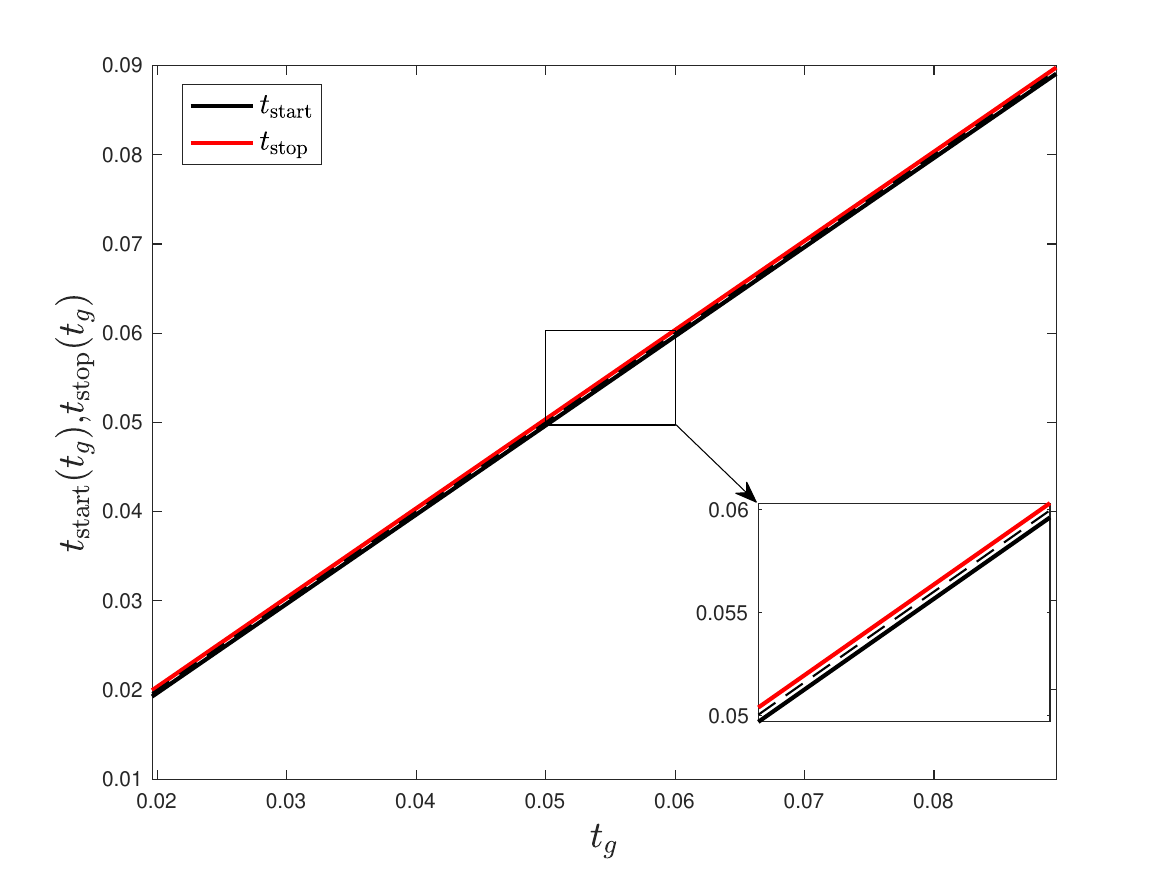}
\label{fig:tstart_tstop_b10}}
\caption{Curves of $t_{\rm start}(t_g)$ (black) and $t_{\rm stop}(t_g)$ (red) versus $t_g$ for different array heights. The dashed line is $t=t_g$.}
\label{fig:tstart_tstop}
\end{figure}

Figs.~\ref{fig:tstart_tstop}\subref{fig:tstart_tstop_b0}
and~\ref{fig:tstart_tstop}\subref{fig:tstart_tstop_b10}
show $t_{\rm start}(t_g)$ (black lines) and
$t_{\rm stop}(t_g)$ (red lines) as functions of $t_g$ for two
values of the array height, namely $b=0$ and $b=10$ m.
For $b=10$ m and the considered parameter values,
$\widehat D_\epsilon$ is much larger than
$\widehat C_\epsilon$, so that the second term in
\eqref{eq:Ellipse_t_csi} has a negligible effect on the
endpoints. Consequently, the interval length
$t_{\rm stop}(t_g)-t_{\rm start}(t_g)$ is practically
independent of $t_g$ and is very close, but not exactly equal,
to $2\widehat C_\epsilon$. For $b=0$, by contrast, the equality
$t_{\rm stop}(t_g)-t_{\rm start}(t_g)=2\widehat D_\epsilon$
holds exactly whenever $t_g\ge\widehat D_\epsilon$. We have also verified numerically that, for the considered
parameter values, the simplified uniform grid described below
provides MNCASE and ANCASE values virtually identical to
those obtained with the grid constructed according to
Theorem~\ref{th:TheoremVULA}:
\[
t_{g_k}=t_{g_0}+k\Delta
\qquad k=0,\ldots,K-1
\]
where $\Delta=2\widehat D_\epsilon$ for $b=0$ and
$\Delta=2\widehat C_\epsilon$ for $b>0$,
$t_{g_0}=t_{\min}+\Delta/2$, and
\[
K=
\left\lceil
\frac{t_{\max}-t_{\min}}{\Delta}
\right\rceil.
\]
Equivalently, in the radial domain, the grid points satisfy
\begin{equation}
\label{eq:UniformRadialGrid}
R_{g_k}
=
R_{g_0}\left(1+k\Delta R_{g_0}\right)^{-1}
\qquad k=0,\ldots,K-1.
\end{equation}
The resulting parameter grid is
$\mathcal G'=\{t_{g_k}:k=0,\ldots,K-1\}$.
For this simplified construction only, the restriction
$\mathcal G'\subseteq\mathcal R'$ is relaxed: the last center is
allowed to lie beyond $t_{\max}$ by less than $\Delta/2$.
The corresponding physical grid is
$\mathcal G=\{\mathbf g_k:k=0,\ldots,K-1\}$, where the grid points
on the selected ray $\varphi_g=0$ have coordinates
\begin{equation}
\label{eq:GridVULA}
\mathbf{g}_k = (\rho_{g_k},0,-b) \quad \quad k=0,\ldots, K-1
\end{equation}
with $\rho_{g_k} = \sqrt{R_{g_k}^2 - b^2}$.

For $b>0$, the accuracy of the uniform-spacing approximation can be
quantified within the adopted surrogate. Since
$|\xi'(t)|=|1-3b^2t^2|\le2$ on $[0,1/b]$, the mean value theorem gives
\begin{equation}
\label{eq:VULAWidthBounds}
\frac{(t-t_g)^2}{\widehat C_\epsilon^2}
\le\Phi(t;t_g)
\le\frac{(t-t_g)^2}{\widehat C_\epsilon^2}
\left[1+4\left(\frac{\widehat C_\epsilon}{\widehat D_\epsilon}\right)^2\right].
\end{equation}
An endpoint not clipped at $0$ or $1/b$ therefore has distance from
$t_g$ between
$\widehat C_\epsilon/\sqrt{1+4(\widehat C_\epsilon/\widehat D_\epsilon)^2}$
and $\widehat C_\epsilon$. This explains why the interval widths are
nearly constant when $\widehat C_\epsilon/\widehat D_\epsilon$ is small.
However, spacing $2\widehat C_\epsilon$ is an approximation and does
not itself certify coverage. Theorem~\ref{th:TheoremVULA} supplies the
optimal interval cover used in the reported simulations.

\begin{figure}[!t]
\centering
\includegraphics[width=0.8\columnwidth]{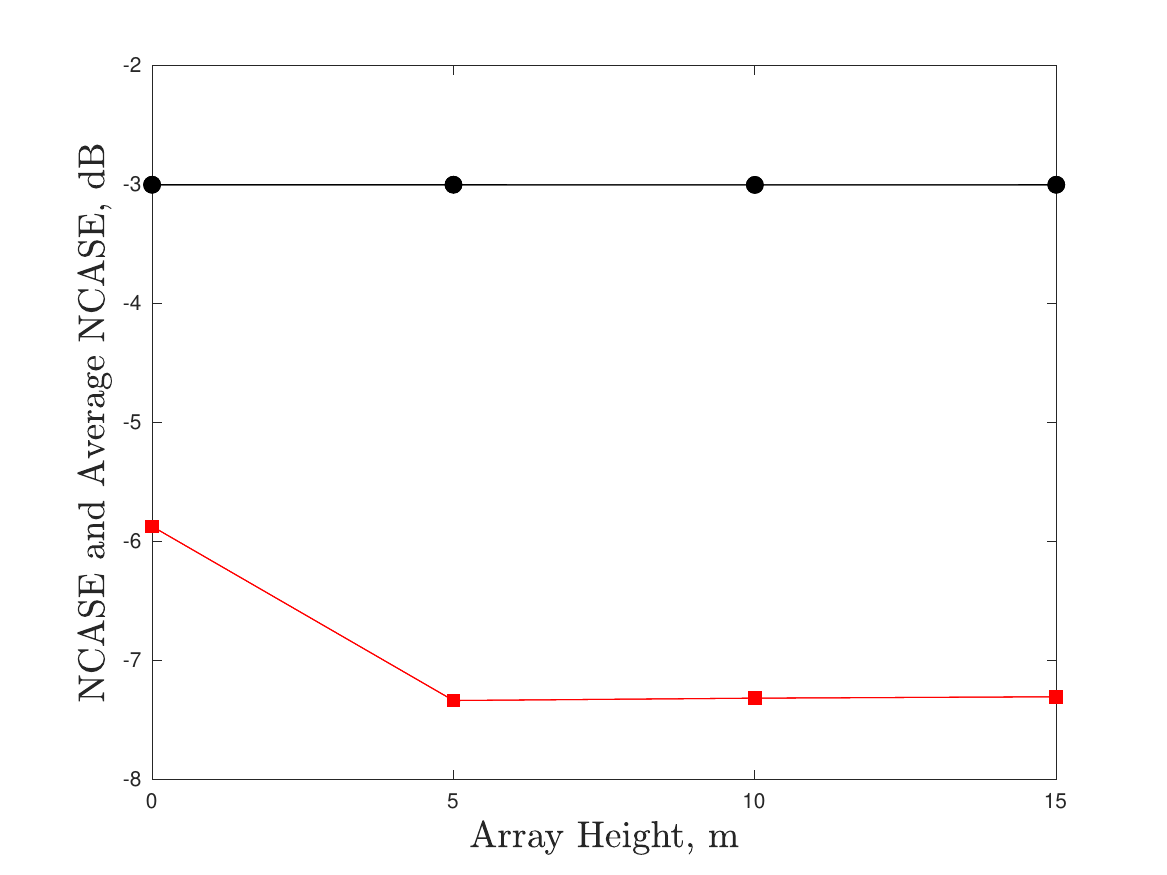}
\caption{MNCASE (black) and ANCASE (red) versus array height for $10^6$ UE positions uniformly distributed with respect to area over the RoI.}
\label{fig:NCASE_VULA}
\end{figure}

The results reported in Fig.~\ref{fig:NCASE_VULA}, which shows MNCASE and ANCASE versus $b$, have been obtained
using the grid constructed according to Theorem~\ref{th:TheoremVULA}. To validate this design, we
have generated $10^6$ UE positions uniformly with respect to area
over the RoI, using the same parameters as in
Fig.~\ref{fig:tstart_tstop} and $10\log_{10}\epsilon=-3$ dB.
Although the grid construction is based on the approximations introduced
above, the NCASE values have been evaluated using the channel model in
\eqref{eq:channel_model}, with exact antenna-to-UE distances and
element-dependent amplitudes. As can be seen, the empirical
MNCASE remains close to the target, while ANCASE is larger for $b=0$ than for
the other considered heights.

Both grid constructions are performed offline. For the HULA, the
implementation evaluates $N'N''$ candidate lattice placements and retains
the one giving the smallest number of selected disks. For the VULA, each
greedy step solves a one-dimensional endpoint problem; under the condition
in Appendix~\ref{App:SolutionVULA}, this reduces to a scalar polynomial
equation for $b>0$ or a closed-form calculation for $b=0$. Once the grid
has been obtained, the associated dictionary is available for repeated
channel-estimation operations. The online estimator therefore uses the
precomputed atoms without repeating the geometric construction.

\section{Channel Estimation Scheme}

We evaluate the dictionaries with the P-SOMP estimator of
\cite{CuiDai_TCOM2022} under the following hybrid receiver model.

\subsection{Signal Model}

We consider a fully connected hybrid receiver in which each of the
$M$ antennas is connected to each of the $N_{\rm RF}\ll M$ RF chains
through a phase-shifter network. During uplink training, the users are
assigned orthogonal resources, which may be separated in time, frequency,
or code. Inter-user interference is therefore absent from the training
observations, and we focus on one arbitrary user~\cite{CuiDai_TCOM2022}.
Let $N_r$ denote the number of time-frequency resources assigned to
that user. The array observation $\mathbf x_i\in\mathbb C^M$ on
resource $i$ is

\begin{equation}
\label{eq: received_signal_digital}
\mathbf{x}_i = \sqrt{p} \phi_i \mathbf{h} + \boldsymbol{\nu}_i \quad i=1,2,\ldots,N_r
\end{equation}

where $p$ is the transmit power, $\phi_i$ is a unit-modulus pilot,
and $\mathbf h$ is the channel in~\eqref{eq:channel}, assumed constant
over the training resources. The additive noise is a circularly symmetric
complex Gaussian vector with distribution
$\boldsymbol\nu_i\sim\mathcal{CN}(\mathbf0,\sigma^2\mathbf I_M)$.
An analog combiner $\mathbf q_{i,c}\in\mathbb C^M$ combines the antenna
signals for RF chain $c$ on resource $i$. Its entries are $\pm1/\sqrt M$,
and different combiners are used across resources and RF chains. The
resulting vector of $N_{\rm RF}$ observations is

\begin{equation}
\label{eq: received_signal_i}
\mathbf{y}_i = \sqrt{p} \phi_i \mathbf{Q}^{\Ttran}_i\mathbf{h} + \mathbf{w}_i \quad i=1,2,\ldots,N_r
\end{equation} 

where $\mathbf Q_i=[\mathbf q_{i,1}\ \cdots\ \mathbf q_{i,N_{\rm RF}}]$
and $\mathbf w_i=\mathbf Q_i^{\Ttran}\boldsymbol\nu_i$.
Multiplication by $\phi_i^*$ gives

\begin{equation}
\label{eq: received_signal_i_2}
\mathbf{z}_i = \sqrt{p}  \mathbf{Q}^{\Ttran}_i\mathbf{h} + {\bf{n}_i} \quad i=1,2,\ldots,N_r
\end{equation}

with $\mathbf n_i=\phi_i^*\mathbf w_i$.
Stacking $\mathbf z=[\mathbf z_1^{\Ttran}\ \cdots\ \mathbf z_{N_r}^{\Ttran}]^{\Ttran}$
yields

\begin{equation}
\label{eq:signal_z}
\mathbf{z} = \sqrt{p} \mathbf{Q}^{\Ttran}\mathbf{h} + {\bf{n}}
\end{equation}  
where $\mathbf{Q} = [\mathbf{Q}_1 \mathbf{Q}_2 \cdots \mathbf{Q}_{N_r}]$ and $\mathbf{n} = [\mathbf{n}^{\Ttran}_1 \mathbf{n}^{\Ttran}_2 \cdots \mathbf{n}^{\Ttran}_{N_r}]^{\Ttran}$.

\subsection{P-SOMP Algorithm}
The P-SOMP algorithm estimates the channel $\mathbf h$ using the
observation vector $\mathbf z$ in \eqref{eq:signal_z}. It relies on the sparse
approximation
\[
\mathbf h \simeq \mathbf W\mathbf h^{\mathcal P}
\]
where $\mathbf W\in\mathbb C^{M\times N_{\rm G}}$ is the dictionary and
$N_{\rm G}=|\mathcal G'|$. Its columns are the steering-vector atoms
generated by the selected parameter grid $\mathcal G'$. For the HULA,
they are given by \eqref{eq:SteeringVectorApp}; for the VULA, they are
associated with the physical points of $\mathcal G$ corresponding to the
selected reciprocal distances.

Under this sparse approximation, whitening the observation vector gives
\begin{equation}
\label{eq:z_white}
\mathbf z_{\rm w}
=
\mathbf D^{-1}\mathbf z
\simeq
\boldsymbol\Psi\mathbf h^{\mathcal P}
+
\mathbf n_{\rm w}
\end{equation}
where $\mathbf{D}$ is the lower triangular factor in the Cholesky decomposition of the correlation matrix $\mathbf{C}$ of $\mathbf{n}$, i.e., $\mathbf{C} = \sigma^2 \mathbf{D} \mathbf{D}^{\Her}$, $\boldsymbol{\Psi} = \sqrt{p}\, \mathbf{D}^{-1} \mathbf{Q}^{\Ttran} \mathbf{W}$, and $\mathbb{E}\left\{\mathbf{n}_{\rm w}\mathbf{n}^{\Her}_{\rm w} \right\} = \sigma^2 \mathbf{I}_{N_rN_{\rm RF}}$. The next step is the computation of vector $\boldsymbol{\gamma} = \boldsymbol{\Psi}^{\Her} \mathbf{z}_{\rm w} \in \mathbb{C}^{N_{\rm G}}$ and the index $\hat g$ of its element with maximum power, i.e.,
\begin{equation}
\label{eq:q_star}
\hat g = \arg \underset{g=1,\ldots,N_{\rm G}}{\max} \left| [\boldsymbol{\gamma}]_g \right|^2
\end{equation}
where $[\boldsymbol{\gamma}]_g$ is the $g$th entry of $\boldsymbol{\gamma}$. According to the P-SOMP algorithm, the estimate ${\kern - 4 pt \hat{\kern 4 pt \mathbf{h}}}$ is finally obtained as
\begin{equation}
\label{eq:estimatePSOMP}
{\kern - 4 pt \hat{\kern 4 pt \mathbf{h}}
=
\hat\alpha\,\hat{\mathbf w}},
\qquad
\hat\alpha
=
(\hat{\boldsymbol\psi}^{\mathsf H}\hat{\boldsymbol\psi})^{-1}
\hat{\boldsymbol\psi}^{\mathsf H}\mathbf z_{\rm w}
\end{equation}
where $\hat{\boldsymbol{\psi}}$ and $\hat{\mathbf{w}}$ are the columns $\hat g$ of $\boldsymbol{\Psi}$ and $\bf W$.  Since we are considering single-path LoS channels, the P-SOMP algorithm stops after the first iteration. A multi-iteration algorithm is described in \cite{CuiDai_TCOM2022} for the case of multipath channels.

The link with the grid-design criterion is that the resulting
estimate is a scaled dictionary atom. Hence, with NCASE evaluated
over the same dictionary $\mathbf W$ and the exact channel
$\mathbf h(\mathbf p)$,
\begin{equation}
\label{eq:EstimationApproximationBound}
\frac{\|\widehat{\mathbf h}-\mathbf h(\mathbf p)\|^2}
{\|\mathbf h(\mathbf p)\|^2}
\ge\operatorname{NCASE}(\mathbf p).
\end{equation}
NCASE characterizes the best available one-atom representation,
which the covering design targets through the adopted approximations.
The attained estimation error also depends on the
compressed observations and noise, and need not equal NCASE even
at high SNR.

\section{Simulation Results with the P-SOMP Algorithm}
We now assess the estimation accuracy of P-SOMP when using the proposed grid design. The performance is measured through the (normalized) mean squared estimation error (MSEE) and the (normalized) maximum squared estimation error (XSEE), defined, respectively, by
\begin{equation}
\label{eq:MSEE}
{\rm MSEE} = \mathbb{E}\left\{\dfrac{\|{\kern - 4 pt \hat{\kern 4 pt \bf h}} - {\bf h}\|^2}{\|{\bf h}\| ^2} \right\}
\end{equation}
and 
\begin{equation}
\label{eq:XSEE}
{\rm XSEE} = \max\left\{\dfrac{\|{\kern - 4 pt \hat{\kern 4 pt \bf h}} - {\bf h}\|^2}{\|{\bf h}\| ^2} \right\}
\end{equation}
where both the expectation and the maximum are the empirical values taken over the simulated channel and noise realizations.

Both ${\rm MSEE}$ and ${\rm XSEE}$ have been computed as a function of the number of pilot symbols, $N_{r}$, the number of RF chains, $N_{\rm RF}$, and the signal-to-noise ratio (SNR). The latter is defined as the ratio between the mean power \textit{at each antenna} and the noise power $\sigma^2$, \textit{for a UE at the edge of the cell}. More precisely, taking \eqref{eq:channel_model} and \eqref{eq: received_signal_digital} into account, and assuming{\footnote {According to the Friis' formula for the free-space path loss.}} $\sqrt{\xi_m} = (4 \pi R_m/\lambda)^{-1}$, where $R_m$ is the distance of the UE from antenna $m$, $p =1$, and $|\phi_i|^2 =1$, we have set
\begin{equation}
\label{eq:SNR}
{\rm SNR} = \dfrac{1}{(4 \pi R_{\max}/\lambda)^2}\dfrac{1}{\sigma^2}
\end{equation}
$R_{\max}$ being the distance of a UE at the cell edge from the center of the array (we are neglecting the dependence of the path loss on the array element). For each SNR value and receiver configuration, the empirical
MSEE and XSEE are evaluated over \(N_{\rm MC}=10^6\)
independent channel and noise realizations.
We consider a horizontal ULA, and a line-of-sight (LoS) propagation scenario. The extension to multipath channels can be carried out as in \cite{CuiDai_TCOM2022}. The system parameters are listed in Table~\ref{tab:SystemParameters}, with an array height of $b=0$ or $b=10$ m. These two settings correspond to the grids reported in Figs.~\ref{fig:GridPointsRoI}\subref{fig:GP0}
and~\ref{fig:GridPointsRoI}\subref{fig:GP10}. As a benchmark, we consider a fully digital system whose number of RF chains matches the number of antennas and which employs least-squares (LS) for channel estimation. For comparison, we generate the reference dictionaries following Algorithm~1 in \cite{CuiDai_TCOM2022}, with sampling parameters chosen to obtain cardinalities comparable to those of the proposed dictionaries.

\begin{figure}[!t]
\centering
\subfloat[$b=0$.\label{fig:MSEb0Np32Nrf5}]{%
\includegraphics[width=0.8\columnwidth]{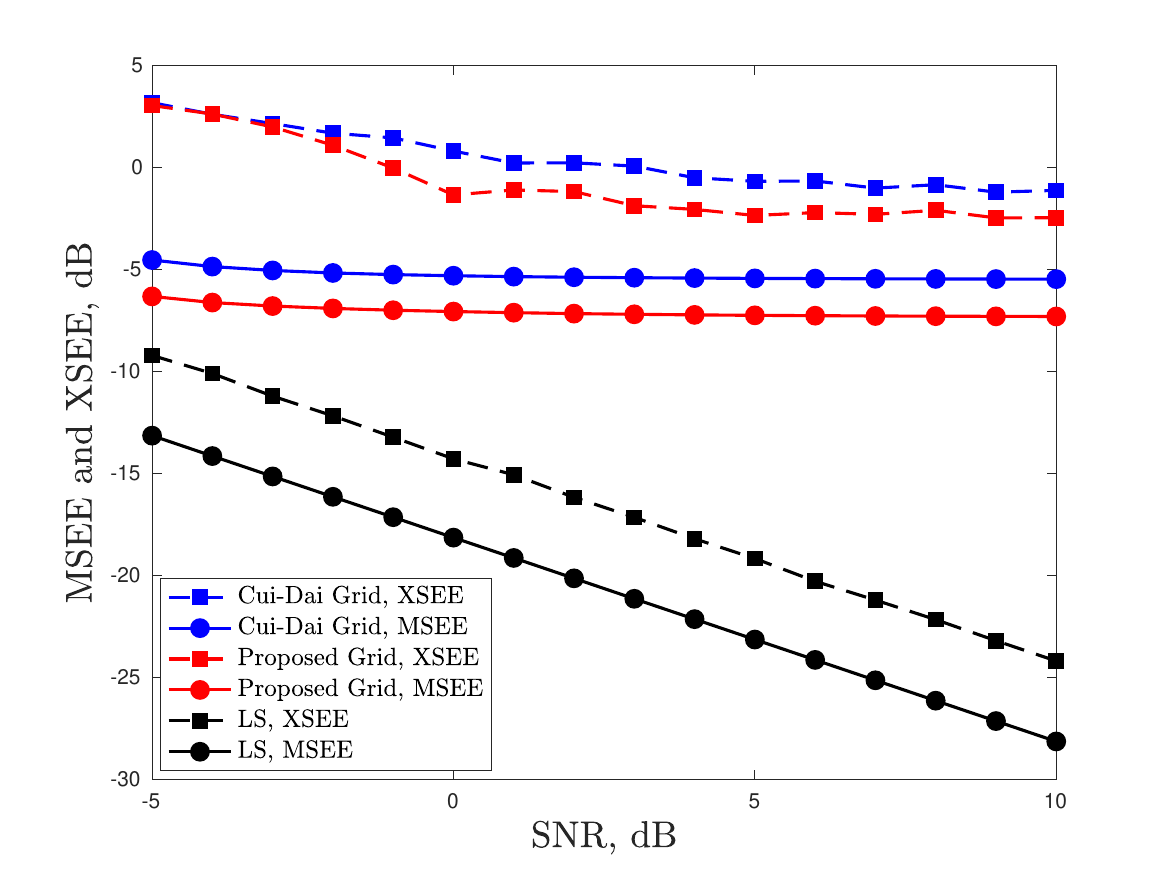}}
\par
\subfloat[$b=10$ m.\label{fig:MSEb10Np32Nrf5}]{%
\includegraphics[width=0.8\columnwidth]{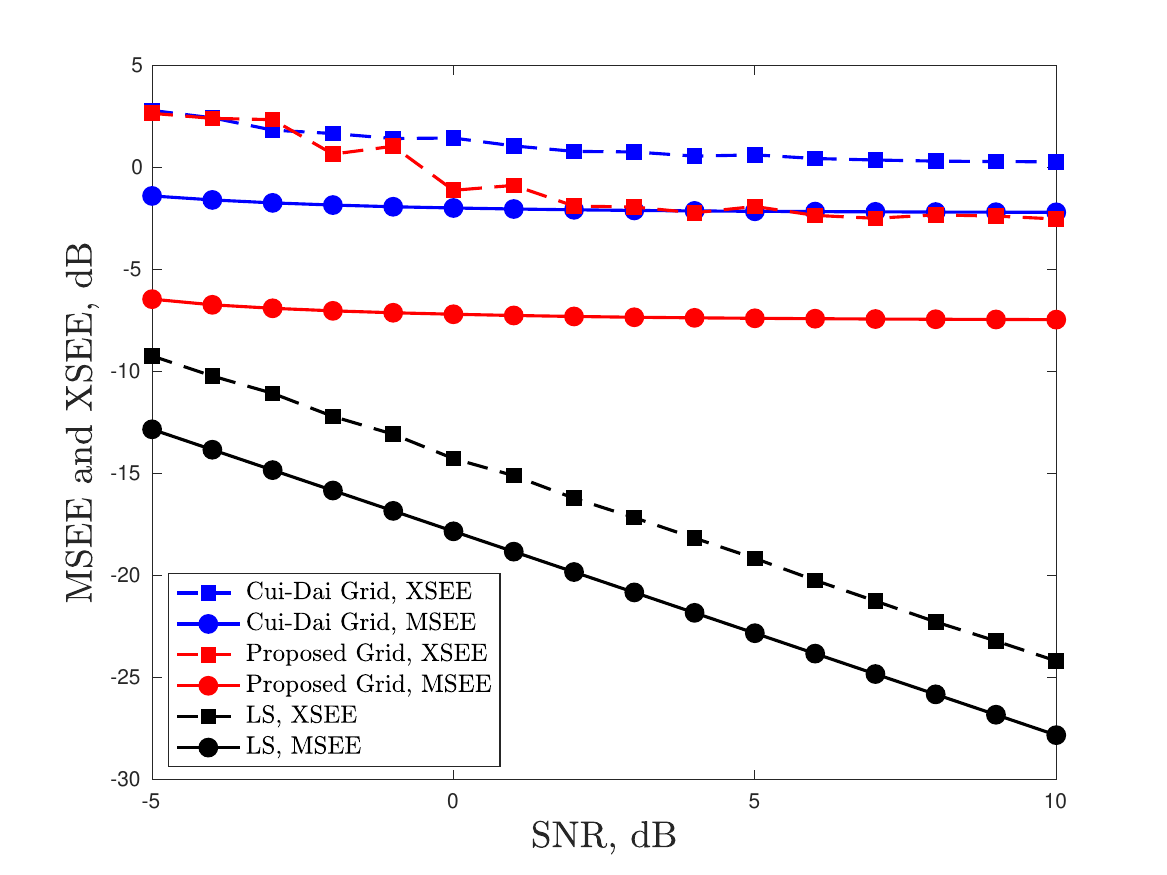}}
\caption{MSEE and XSEE versus SNR for $b=0$ and $b=10$ m.}
\label{fig:MSE_PSOMP_Np32Nrf5}
\end{figure}

Fig.~\ref{fig:MSE_PSOMP_Np32Nrf5} shows ${\rm MSEE}$ and ${\rm XSEE}$ as a function of $\mathrm{SNR}$ for $N_{\rm RF} = 5$ and  $N_{r} = 32$. As can be seen, the proposed grid outperforms the design in \cite{CuiDai_TCOM2022} in terms of both ${\rm MSEE}$ and ${\rm XSEE}$. In particular, the gap in ${\rm MSEE}$ is approximately $2$ dB for $b=0$ and  exceeds $5$ dB when $b=10$ m. Moreover, the performance of the proposed design is essentially independent of the array height. As expected, the fully digital estimator significantly outperforms the grid-based estimators designed for hybrid architectures. Naturally, this improvement comes at the cost of increased hardware complexity, which may be unsustainable in many practical scenarios.  

\begin{figure}[!t]
\centering
\subfloat[$N_{\rm RF}=10$, $N_r=16$.\label{fig:Np16Nrf10}]{%
\includegraphics[width=0.8\columnwidth]{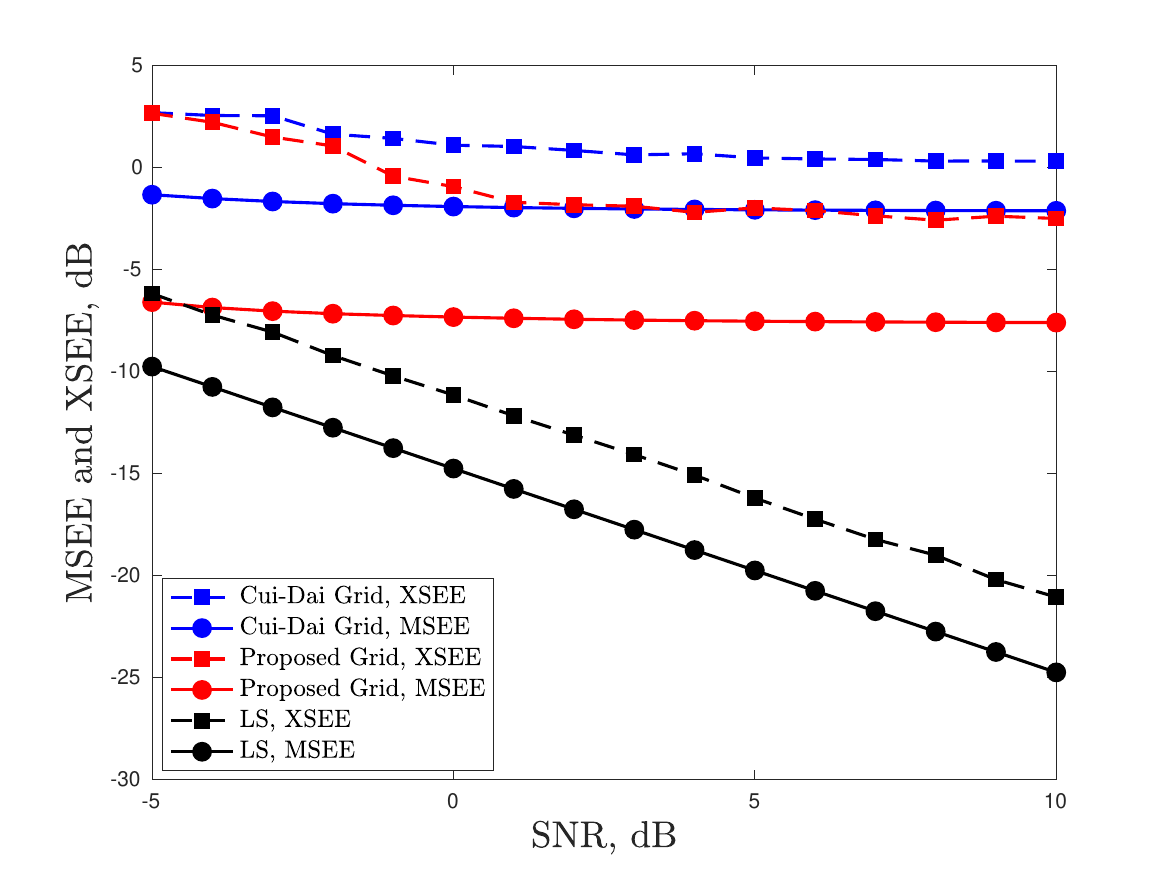}}
\par
\subfloat[$N_{\rm RF}=20$, $N_r=8$.\label{fig:Np8Nrf20}]{%
\includegraphics[width=0.8\columnwidth]{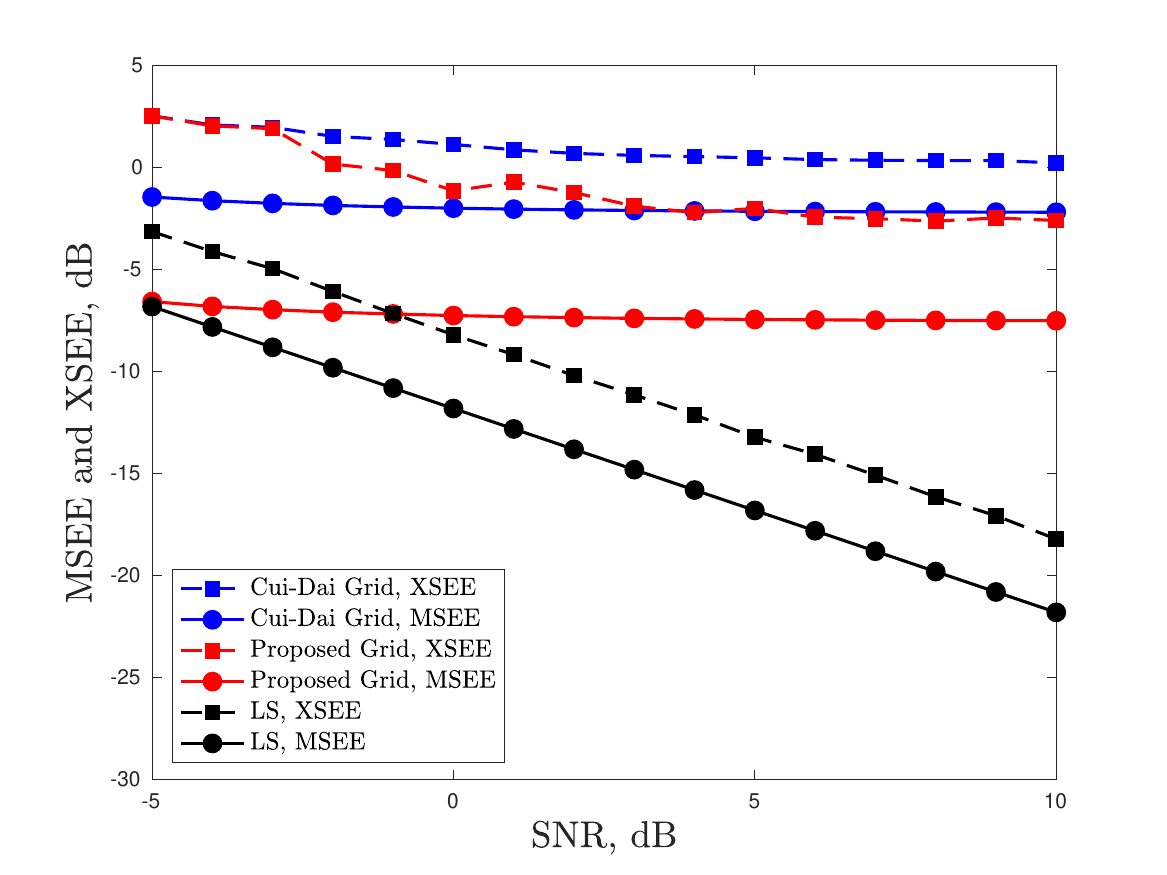}}
\caption{MSEE and XSEE versus SNR for $b=10$ m and two values of $(N_{\rm RF},N_r)$.}
\label{fig:MSE_PSOMP_DifferentNpNrf}
\end{figure}

Fig.~\ref{fig:MSE_PSOMP_DifferentNpNrf} illustrates ${\rm MSEE}$ and ${\rm XSEE}$ versus $\mathrm{SNR}$ for $b = 10$ m and two different values of the pair $(N_{\rm RF},N_{r})$. For the considered configurations, which have the same product $N_{\rm RF}N_r=160$, the grid-based estimators exhibit nearly
identical performance. This indicates that, in the considered
grid-limited regime, redistributing a fixed number of measurements
between RF chains and pilot resources has little effect. 
Conversely, as expected, the performance of the fully digital LS estimator degrades when the number of pilot symbols is reduced.

\section{Conclusion and Future Work}
This paper proposed a deterministic near-field grid design for ULAs and hybrid
receivers, linking grid density directly to a prescribed NCASE target. For a
HULA, the problem becomes a transformed-domain disk covering addressed through
a low-cardinality hexagonal-lattice construction; for a VULA, it becomes a
one-dimensional interval-covering problem solved optimally by a greedy
farthest-reaching procedure. For the considered configurations, numerical evaluation using the exact
spherical-wave channel model yields empirical maximum approximation errors
close to the prescribed target. The HULA simulations also show improved
single-path P-SOMP estimation relative to the tested polar-domain
dictionaries of comparable cardinality. Future work will consider multipath propagation,
more general arrays, refined finite-region coverings, and off-grid refinement.

\appendices

\section{Approximation of the Admissible Region}
\label{App:AdmissRegion}

To motivate the elliptical approximation, expand the finite-array
correlation in~\eqref{eq: Correlation1} around its maximum at $(0,0)$.
The first derivatives and the mixed second derivative vanish there.
The remaining second derivatives are
\begin{equation}
\label{eq:AppA_1}
\begin{aligned}
\left.\frac{\partial^2\Lambda}{\partial A^2}\right|_{(0,0)}
&=-\frac{M^2-1}{6}\\
\left.\frac{\partial^2\Lambda}{\partial B^2}\right|_{(0,0)}
&=-\frac{(M^2-1)(M^2-4)}{90}.
\end{aligned}
\end{equation}
The second-order Taylor approximation is therefore

\begin{equation}
\label{eq:AppA_2}
\Lambda(A,B) \approx 1 - \left[\dfrac{M^2-1}{12} A^2 +\dfrac{(M^2-4)(M^2 -1)}{180} B^2\right] \raisebox{2 pt}{.}
\end{equation}
Based on \eqref{eq:AppA_2}, the condition $\Lambda(A,B) \ge 1 -\epsilon$ can be written as 
\begin{equation}
\label{eq:AppA_3}
\dfrac{A^2}{\tilde A^2_{\epsilon}}  +\dfrac{B^2}{\tilde B^2_{\epsilon}}  \le 1
\end{equation}
where
\begin{equation}
\label{eq:AppA_4}
\tilde A^2_{\epsilon} = \dfrac{12 \epsilon}{M^2 -1} \quad \quad \tilde B^2_{\epsilon} = \dfrac{180 \epsilon}{(M^2 - 4)(M^2 - 1)}\raisebox{2 pt}{.}
\end{equation}

The quadratic coefficients are the variances of the centered antenna
index $m$ and its square $m^2$, respectively, with uniform weights
over the $M$ antennas. Symmetry makes the odd moments vanish and
eliminates the mixed term. The local quadratic model therefore
explains the elliptical shape and its unequal axes.

The calibrated ellipse~\eqref{eq:Ellipse} uses the first positive
axis intercepts of the finite-array correlation level
$\Lambda=1-\epsilon$, rather than the Taylor intercepts.
Both constructions coincide to leading order as $\epsilon\to0$.
In particular,
\begin{equation}
\label{eq:EllipseAxisRatioLimit}
\lim_{\epsilon\to0}
\frac{\widehat A_\epsilon}{\widehat B_\epsilon}
=\sqrt{\frac{M^2-4}{15}}.
\end{equation}
For $M=255$, this limit is approximately $65.84$, consistent with
Table~\ref{tab:table0}. The numerical calibration retains the exact
axis crossings at the chosen target even when the small-error
Taylor approximation is less accurate. The angular intercept solves

\begin{equation}
\label{eq:AngularBoundaryRoot}
\left[\dfrac{\sin{(MA/2})}{M\sin{(A/2)}}\right]^2 = 1 - \epsilon
\end{equation}
whereas $\widehat{B}_{\epsilon}$ is obtained numerically as the
smallest positive solution of
\[
\left|
\frac{1}{M}
\sum_{m=-(M-1)/2}^{(M-1)/2} e^{jBm^2}
\right|^2 = 1-\epsilon.
\]
It turns out that $\widehat{A}_{\epsilon}$ and $\widehat{B}_{\epsilon}$ are slightly larger than $\tilde A_{\epsilon}$ and $\tilde B_{\epsilon}$, respectively, and that the elliptical region delimited by \eqref{eq:Ellipse} provides a better approximation of the admissible region than that delimited by \eqref{eq:AppA_3}.

\section{Hexagonal-Lattice Covering Construction}
\label{App:Solution}

Using the scaling in Section~II, each ellipse~\eqref{eq:EllipseGEta}
becomes the unit disk
\begin{equation}
\label{eq:AppC3}
\mathsf C_g:
(\tilde\Gamma-\tilde\Gamma_g)^2+
(\tilde\eta-\tilde\eta_g)^2\le1.
\end{equation}
It remains to cover $\mathcal R''$ with a finite set of these disks.

We restrict the centers to a hexagonal lattice. Such lattice coverings are
classical in the geometry of numbers and sphere-covering theory
\cite{ConwaySloane1999}, and the hexagonal arrangement is asymptotically
optimal for planar circle covering as the covered region becomes large
\cite{Kershner_1939}. Its
points are
\begin{equation}
\label{eq:HexagonalLattice}
\left[
\begin{array}{c}
\tilde\Gamma_{m,n}\\
\tilde\eta_{m,n}
\end{array}
\right]
=
\left[
\begin{array}{c}
\tilde\Gamma_0\\
\tilde\eta_0
\end{array}
\right]
+
\kappa
\left[
\begin{array}{cc}
\cos\omega&-\sin\omega\\
\sin\omega&\cos\omega
\end{array}
\right]
(m\mathbf u_1+n\mathbf u_2)
\end{equation}
where $\mathbf u_1=[0\ \ 1]^{\Ttran}$,
$\mathbf u_2=[\sqrt3/2\ \ 1/2]^{\Ttran}$, $m,n\in\mathbb Z$, and
$\kappa=\sqrt3$ for unit covering disks. By symmetry it is sufficient to use
$\omega\in[0,\pi/3)$.

The nearest-neighbor distance is $\kappa=\sqrt3$, and the Voronoi
cell of each lattice point is a regular hexagon with circumradius one.
The unit disk centered at that point contains its entire cell. Since
the Voronoi cells tile the plane, the associated disks cover the plane
without gaps. For the bounded region $\mathcal R''$, only finitely many
of these disks are needed. The following selection rule retains all
disks intersecting its interior; boundary-dependent changes in this
finite set explain the role of lattice translation and orientation.

For a fixed triplet $(\tilde\Gamma_0,\tilde\eta_0,\omega)$, let
$\mathcal L$ denote the lattice in \eqref{eq:HexagonalLattice},
$\mathcal C=\{\mathsf C_{m,n}:m,n\in\mathbb Z\}$ the associated unit disks,
and select
\[
\overline{\mathcal C}
=
\{\mathsf C\in\mathcal C:
\mathsf C\cap\operatorname{int}(\mathcal R'')\ne\emptyset\}
\]
where $\operatorname{int}(\mathcal R'')$ denotes the \textit{interior} of
$\mathcal R''$. Let $\overline{\mathcal L}$ be the set of the centers of
the disks in $\overline{\mathcal C}$. This rule excludes disks that
intersect $\mathcal R''$ only on its boundary. Since $\mathcal R''$ is
bounded and the lattice is discrete, $\overline{\mathcal C}$ is finite.
Moreover, the selected disks cover $\operatorname{int}(\mathcal R'')$.
Since their finite union is closed, they also cover
$\mathcal R''=\operatorname{cl}(\operatorname{int}(\mathcal R''))$,
where $\operatorname{cl}(\cdot)$ denotes the \textit{closure} of its
argument. Therefore, the restricted lattice-design problem can be formalized as
\begin{equation}
\label{eq:AppC6}
\begin{aligned}
\underset{\tilde\Gamma_0,\tilde\eta_0,\omega}{\min}\quad
&|\overline{\mathcal L}|\\
\mathrm{s.t.}\quad
&\bigcup_{\mathsf C_{m,n}\in\overline{\mathcal C}}
\mathsf C_{m,n}\supseteq\mathcal R''.
\end{aligned}
\end{equation}

The intersection rule fixes the selected disks for each placement;
only translation and orientation are optimized.

Now, observe that, for each orientation, translations need only be represented within one
fundamental hexagonal cell. For $\omega=0$ we use
\[
\begin{aligned}
\mathcal H_0=\{(\tilde\Gamma,\tilde\eta)\in\mathbb R^2:
&\ |\tilde\eta|\le\sqrt3/2,\\
&\ |\sqrt3\tilde\Gamma+\tilde\eta|\le\sqrt3,\\
&\ |\tilde\eta-\sqrt3\tilde\Gamma|\le\sqrt3\}
\end{aligned}
\]

and use its rotation $\mathcal H_0^\omega$ for $\omega\ne0$.
We retain the tested placement with the smallest
$|\overline{\mathcal L}|$ and recover the parameter grid using
\eqref{eq:HULAGridCenters}.

\section{Admissible Intervals for the Vertical ULA}
\label{App:SolutionVULA}

For $b>0$, substitute $\bar t=bt$ and $\bar t_g=bt_g$ in
\eqref{eq:Itg}--\eqref{eq:csi_t}. The admissibility condition becomes
\begin{equation}
\label{eq:system2}
F(\bar t; \bar t_g) \overset{\mathrm{def}}{=} \dfrac{(\bar t-\bar t_g)^2}{\mathrm{C}_{\epsilon}^2}+\dfrac{[\psi(\bar t)-\psi_g]^2}{\mathrm{D}_{\epsilon}^2} - 1 \le 0
\end{equation}
where $0\le\bar t\le1$, $0<\bar t_g\le1$,
$\psi(\bar t)=\bar t(1-\bar t^2)$, $\psi_g=\psi(\bar t_g)$,
$\mathrm C_\epsilon=b\widehat C_\epsilon$, and
$\mathrm D_\epsilon=b\widehat D_\epsilon$.
We show that the admissible set is an interval if
\begin{equation}
\label{eq:ratio_Height_delta}
\dfrac{b}{\delta} > \dfrac{1}{2\sqrt{3}}\dfrac{\widehat{A}_{\epsilon}}{\widehat{B}_{\epsilon}}
\end{equation}
which is a mild height-to-spacing condition. Differentiation gives
\begin{equation}
\label{eq:derLambda}
F'(\bar t; \bar t_g) = \dfrac{2(\bar t-\bar t_g)}{\mathrm{C}_{\epsilon}^2}\left[ 1 + \dfrac{\mathrm{C}_{\epsilon}^2}{\mathrm{D}_{\epsilon}^2} \, g(\bar t; \bar t_g)(1 - 3 {\bar t}^{\,2})\right]
\end{equation}
where $g(\bar t;\bar t_g)=1-\bar t^2-\bar t_g\bar t-\bar t_g^2$.
The bound needed below can be established explicitly. Write
$u=\bar t$, $v=\bar t_g$, and
$q(u)=\min_{0\le v\le1}g(u;v)(1-3u^2)$.
Monotonicity of $g(u;v)$ in $v$ gives
\[
q(u)=
\begin{cases}
-u(1+u)(1-3u^2) & 0\le u\le1/\sqrt3\\
(1-u^2)(1-3u^2) & 1/\sqrt3\le u\le1.
\end{cases}
\]
In the first case, $u(1-3u^2)\le2/9$ and
$u^2(1-3u^2)\le1/12$, so the expression is at least
$-11/36>-1/3$. In the second case, it equals
$3(u^2-2/3)^2-1/3\ge-1/3$. Therefore
$g(\bar t;\bar t_g)(1-3\bar t^2)\ge-1/3$ throughout $[0,1]^2$.
Accordingly, if

\begin{equation}
\label{eq:Condition}
{1-\frac{\mathrm C_\epsilon^2}{3\mathrm D_\epsilon^2}>0}
\end{equation}

then
\[
\operatorname{sign}F'(\bar t;\bar t_g)
=\operatorname{sign}(\bar t-\bar t_g).
\]
Thus, $F$ is strictly decreasing to the left of $\bar t_g$ and
strictly increasing to its right, with minimum
$F(\bar t_g;\bar t_g)=-1$. Its nonpositive sublevel set is consequently
a single interval. Substituting $\mathrm C_\epsilon=b\widehat C_\epsilon$
and $\mathrm D_\epsilon=b\widehat D_\epsilon$ from~\eqref{eq:hatChatD}
shows that~\eqref{eq:Condition} is equivalent
to~\eqref{eq:ratio_Height_delta}.

The interval endpoints are determined by the values of $F$ at the
boundaries of $[0,1]$. If $F(0;\bar t_g)\le0$, set
$\bar t_{\rm start}(\bar t_g)=0$; otherwise, use the unique root of
$F(\bar t;\bar t_g)=0$ in $(0,\bar t_g)$.
Similarly, if $F(1;\bar t_g)\le0$, set
$\bar t_{\rm stop}(\bar t_g)=1$; otherwise, use the unique root in
$(\bar t_g,1)$. Continuity, strict monotonicity on each side of the
center, and the negative minimum ensure existence and uniqueness of
these interior roots. These definitions also include roots exactly
at $0$ or $1$. If both boundary values are nonpositive, the entire
interval $[0,1]$ is admissible. Returning to the original coordinates gives

\begin{equation}
\label{eq:admissible_interval}
\mathcal I(t_g)=\left[t_{\rm start}(t_g),t_{\rm stop}(t_g)\right]
\end{equation}
where
\begin{equation}
\label{eq:tstart_tstop_Appendix}
t_{\rm start}(t_g)
=
\frac{1}{b}\,\bar t_{\rm start}(b t_g)
\qquad
 t_{\rm stop}(t_g)
=
\frac{1}{b}\,\bar t_{\rm stop}(b t_g).
\end{equation}

For an interior endpoint $\bar t_*(\bar t_g)$, implicit differentiation
of $F(\bar t_*;\bar t_g)=0$ yields

\begin{equation}
\label{eq:ImplDerivative}
\dfrac{d\bar{t}_{\ast}}{d\bar t_g} = \dfrac{1 + \dfrac{\mathrm{C}_{\epsilon}^2}{\mathrm{D}_{\epsilon}^2} \, g(\bar{t}_{\ast}; \bar t_g)(1 - 3 {\bar t_g}^{\,2})}{1 + \dfrac{\mathrm{C}_{\epsilon}^2}{\mathrm{D}_{\epsilon}^2} \, g(\bar{t}_{\ast}; \bar t_g)(1 - 3 {\bar{t}_{\ast}}^{\,2})} \raisebox{2 pt}{.}
\end{equation}

Under~\eqref{eq:Condition}, numerator and denominator are positive by
the same bound on $g$. Interior endpoints therefore increase with the
center; clipped endpoints are constant. The resulting continuous
endpoints are nondecreasing. Since $F(\bar t_g;\bar t_g)=-1$, each
center also has an admissible neighborhood, as required in
Theorem~\ref{th:TheoremVULA}. For $b=0$, the endpoints reduce to
\begin{equation}
\label{eq:tstart_tstop_b0_App}
t_{\rm start}(t_g) = \max\{0, t_g - \widehat{D}_{\epsilon}\}, \quad t_{\rm stop}(t_g) = t_g + \widehat{D}_{\epsilon}.
\end{equation}

\bibliographystyle{IEEEtran}
\bibliography{mybib}

@inproceedings{Castro2020,
  author      = {Castro, Carlos and Elschner, Robert and Merkle, Thomas and Schubert, Colja and Freund, Ronald},
  title       = {Long-range high-speed {THz}-wireless transmission in the 300 {GHz} band},
  booktitle   = {Proc. 3rd Int. Workshop Mobile Terahertz Syst. (IWMTS)},
  pages       = {1-4},
  year        = {2020},
  doi         = {10.1109/IWMTS49292.2020.9166263}
}

@article{Huang_MM_2011,
  author      = {Huang, Kao-Cheng and Wang, Zhaocheng},
  title       = {Terahertz terabit wireless communication},
  journal     = {IEEE Microw. Mag.},
  volume      = {12},
  number      = {4},
  pages       = {108-116},
  year        = {2011},
  doi         = {10.1109/MMM.2011.940596}
}

@article{Rappaport_2019,
  author      = {Rappaport, Theodore S. and Xing, Yunchou and Kanhere, Ojas and Ju, Shihao and Madanayake, Arjuna and Mandal, Soumyajit and Alkhateeb, Ahmed and Trichopoulos, Georgios C.},
  title       = {Wireless communications and applications above 100 {GHz}: {Opportunities} and challenges for {6G} and beyond},
  journal     = {IEEE Access},
  volume      = {7},
  pages       = {78729-78757},
  year        = {2019},
  doi         = {10.1109/ACCESS.2019.2921522}
}

@article{Ning_OJCS_2023,
  author      = {Ning, Boyu and Tian, Zhongbao and Mei, Weidong and Chen, Zhi and Han, Chong and Li, Shaoqian and Yuan, Jinhong and Zhang, Rui},
  title       = {Beamforming technologies for ultra-massive {MIMO} in terahertz communications},
  journal     = {IEEE Open J. Commun. Soc.},
  volume      = {4},
  pages       = {614-658},
  year        = {2023},
  doi         = {10.1109/OJCOMS.2023.3245669}
}

@article{molisch2017hybrid,
  author      = {Molisch, Andreas F. and Ratnam, Vishnu V. and Han, Shengqian and Li, Zheda and Nguyen, Sinh Le Hong and Li, Linsheng and Haneda, Katsuyuki},
  title       = {Hybrid beamforming for massive {MIMO}: {A} survey},
  journal     = {IEEE Commun. Mag.},
  volume      = {55},
  number      = {9},
  pages       = {134--141},
  month       = sep,
  year        = {2017},
  doi         = {10.1109/MCOM.2017.1600400}
}

@article{heath2016overview,
  author      = {Heath, Robert W. and Gonzalez-Prelcic, Nuria and Rangan, Sundeep and Roh, Wonil and Sayeed, Akbar M.},
  title       = {An overview of signal processing techniques for millimeter wave {MIMO} systems},
  journal     = {IEEE J. Sel. Topics Signal Process.},
  volume      = {10},
  number      = {3},
  pages       = {436--453},
  month       = apr,
  year        = {2016},
  doi         = {10.1109/JSTSP.2016.2523924}
}

@article{el2014spatially,
  author      = {El Ayach, Omar and Rajagopal, Sridhar and Abu-Surra, Shadi and Pi, Zhouyue and Heath, Robert W.},
  title       = {Spatially sparse precoding in millimeter wave {MIMO} systems},
  journal     = {IEEE Trans. Wireless Commun.},
  volume      = {13},
  number      = {3},
  pages       = {1499--1513},
  month       = mar,
  year        = {2014},
  doi         = {10.1109/TWC.2014.011714.130846}
}

@article{huang2020dfthc,
  author      = {Huang, Yu and Liu, Chen and Song, Yunchao and Yu, Xiaolei},
  title       = {{DFT} codebook-based hybrid precoding for multiuser {mmWave} massive {MIMO} systems},
  journal     = {EURASIP J. Adv. Signal Process.},
  volume      = {2020},
  number      = {1},
  year        = {2020},
  note        = {{Art. no. 11}},
  doi         = {10.1186/s13634-020-00669-4}
}

@techreport{3GPP38214,
  author      = {{3rd Generation Partnership Project (3GPP)}},
  title       = {{NR}; physical layer procedures for data},
  number      = {3GPP TS 38.214, Version 19.5.0, Release 19},
  institution = {3rd Generation Partnership Project (3GPP)},
  type        = {Tech. Spec.},
  month       = sep,
  year        = {2026}
}

@article{giordani2019nr,
  author      = {Giordani, Marco and Polese, Michele and Roy, Arnab and Castor, Douglas and Zorzi, Michele},
  title       = {A tutorial on beam management for {3GPP NR} at {mmWave} frequencies},
  journal     = {IEEE Commun. Surveys Tuts.},
  volume      = {21},
  number      = {1},
  pages       = {173-196},
  year        = {2019},
  doi         = {10.1109/COMST.2018.2869411}
}

@article{xiao2016codebook,
  author      = {Xiao, Zhenyu and Xia, Pengfei and Xia, Xiang-Gen},
  title       = {Codebook design for millimeter-wave channel estimation with hybrid precoding structure},
  journal     = {IEEE Trans. Wireless Commun.},
  volume      = {16},
  number      = {1},
  pages       = {141--153},
  month       = jan,
  year        = {2017},
  doi         = {10.1109/TWC.2016.2619705}
}

@article{qi2022hierarchical,
  author      = {Qi, Chenhao and Chen, Kai and Dobre, Octavia A. and Li, Geoffrey Ye},
  title       = {Hierarchical codebook-based multiuser beam training for millimeter wave massive {MIMO}},
  journal     = {IEEE Trans. Wireless Commun.},
  volume      = {19},
  number      = {12},
  pages       = {8142--8152},
  month       = dec,
  year        = {2020},
  doi         = {10.1109/TWC.2020.3019523}
}

@article{chen2018nonuniform,
  author      = {Chen, Yun and Chen, Dongming and Jiang, Tao},
  title       = {Non-uniform quantization codebook-based hybrid precoding to reduce feedback overhead in millimeter wave {MIMO} systems},
  journal     = {IEEE Trans. Commun.},
  volume      = {67},
  number      = {4},
  pages       = {2779--2791},
  month       = apr,
  year        = {2019},
  doi         = {10.1109/TCOMM.2018.2890227}
}

@article{liu2021codebook,
  author      = {Liu, Guo and Deng, Haijun and Yang, Kai and Zhu, Zhenyu and Liu, Jing and Dong, Hui},
  title       = {A new design of codebook for hybrid precoding in millimeter-wave massive {MIMO} systems},
  journal     = {Symmetry},
  volume      = {13},
  number      = {5},
  year        = {2021},
  note        = {{Art. no. 743}},
  doi         = {10.3390/sym13050743}
}

@article{bhuyan2024adaptive,
  author      = {Bhuyan, Mridul and Sarma, Kandarpa Kumar and Misra, Debdeep D. and Guha, Koushik and Iannacci, Jacopo},
  title       = {Adaptive hybrid beamforming codebook design using multi-agent reinforcement learning for multiuser multiple-input--multiple-output systems},
  journal     = {Appl. Sci.},
  volume      = {14},
  number      = {16},
  year        = {2024},
  note        = {{Art. no. 7109}},
  doi         = {10.3390/app14167109}
}

@article{lee2016channel,
  author      = {Lee, Jae and Gil, Gil-Tae and Lee, Yong H.},
  title       = {Channel estimation via orthogonal matching pursuit for hybrid {MIMO} systems in millimeter wave communications},
  journal     = {IEEE Trans. Commun.},
  volume      = {64},
  number      = {6},
  pages       = {2370--2386},
  month       = jun,
  year        = {2016},
  doi         = {10.1109/TCOMM.2016.2557791}
}

@article{tang2019offgrid,
  author      = {Tang, Hui and Wang, Jun and He, Lin},
  title       = {Off-grid sparse {Bayesian} learning-based channel estimation for {mmWave} massive {MIMO} uplink},
  journal     = {IEEE Wireless Commun. Lett.},
  volume      = {8},
  number      = {1},
  pages       = {45--48},
  month       = feb,
  year        = {2019}
}

@article{LuZeng_TWC2022,
  author      = {Lu, Haiquan and Zeng, Yong},
  title       = {Communicating with extremely large-scale array/surface: {Unified} modeling and performance analysis},
  journal     = {IEEE Trans. Wireless Commun.},
  volume      = {21},
  number      = {6},
  pages       = {4039-4053},
  month       = jun,
  year        = {2022},
  doi         = {10.1109/TWC.2021.3126384}
}

@article{KosasihBjornson_TWC2024,
  author      = {Kosasih, Alva and Bj{\"o}rnson, Emil},
  title       = {Finite beam depth analysis for large arrays},
  journal     = {IEEE Trans. Wireless Commun.},
  volume      = {23},
  number      = {8},
  pages       = {10015-10029},
  month       = aug,
  year        = {2024},
  doi         = {10.1109/TWC.2024.3368013}
}

@article{CuiDai_TCOM2022,
  author      = {Cui, Mingyao and Dai, Linglong},
  title       = {Channel estimation for extremely large-scale {MIMO}: {Far}-field or near-field?},
  journal     = {IEEE Trans. Commun.},
  volume      = {70},
  number      = {4},
  pages       = {2663-2677},
  year        = {2022},
  doi         = {10.1109/TCOMM.2022.3146400}
}

@inproceedings{10403506,
  author      = {Demir, {\"O}zlem Tu{\u{g}}fe and Bj{\"o}rnson, Emil},
  title       = {A new polar-domain dictionary design for the near-field region of extremely large aperture arrays},
  booktitle   = {Proc. IEEE 9th Int. Workshop Comput. Adv. Multi-Sensor Adapt. Process. (CAMSAP)},
  pages       = {251-255},
  year        = {2023},
  doi         = {10.1109/CAMSAP58249.2023.10403506}
}

@article{10123941,
  author      = {Wu, Zidong and Dai, Linglong},
  title       = {Multiple access for near-field communications: {SDMA} or {LDMA}?},
  journal     = {IEEE J. Sel. Areas Commun.},
  volume      = {41},
  number      = {6},
  pages       = {1918-1935},
  year        = {2023},
  doi         = {10.1109/JSAC.2023.3275616}
}

@article{zhang2024sparse,
  author      = {Zhang, Xing and Zhang, Haiyang and Eldar, Yonina C.},
  title       = {Near-field sparse channel representation and estimation in {6G} wireless communications},
  journal     = {IEEE Trans. Commun.},
  volume      = {72},
  number      = {1},
  pages       = {450-464},
  month       = jan,
  year        = {2024},
  doi         = {10.1109/TCOMM.2023.3322449}
}

@article{zhang2024codebook,
  author      = {Zhang, Xiangyu and Zhang, Haiyang and Zhang, Jianjun and Li, Chunguo and Huang, Yongming and Yang, Luxi},
  title       = {Codebook design for extremely large-scale {MIMO} systems: {Near}-field and far-field},
  journal     = {IEEE Trans. Commun.},
  volume      = {72},
  number      = {2},
  pages       = {1191-1206},
  month       = feb,
  year        = {2024},
  doi         = {10.1109/TCOMM.2023.3329224}
}

@article{chen2023hierarchical,
  author      = {Chen, Jiawei and Gao, Feifei and Jian, Mengnan and Yuan, Wanmai},
  title       = {Hierarchical codebook design for near-field {mmWave} {MIMO} communications systems},
  journal     = {IEEE Wireless Commun. Lett.},
  volume      = {12},
  number      = {11},
  pages       = {1926--1930},
  month       = nov,
  year        = {2023}
}

@article{zhang2022fast,
  author      = {Zhang, Yunpu and Wu, Xun and You, Changsheng},
  title       = {Fast near-field beam training for extremely large-scale array},
  journal     = {IEEE Wireless Commun. Lett.},
  volume      = {11},
  number      = {12},
  pages       = {2625--2629},
  month       = dec,
  year        = {2022},
  doi         = {10.1109/LWC.2022.3212344}
}

@inproceedings{zhu2025clusterbased,
  author      = {Zhu, Jikun and Ma, Xiaochen and Wang, Zheng and Huang, Yongming},
  title       = {Cluster-based low-complexity codebook design for hierarchical beam training in {XL-MIMO}},
  booktitle   = {Proc. Int. Wireless Commun. Mobile Comput. (IWCMC)},
  pages       = {836-840},
  year        = {2025},
  doi         = {10.1109/IWCMC65282.2025.11059671}
}

@article{Zheng_TCOM2025,
  author      = {Zheng, Feng and Yu, Hongkang and Wang, Chenchen and Sun, Luyang and Wu, Qingqing and Chen, Yijian},
  title       = {Extremely large-scale array systems: {Near}-field codebook design and performance analysis},
  journal     = {IEEE Trans. Commun.},
  volume      = {73},
  number      = {8},
  pages       = {6798-6812},
  year        = {2025},
  doi         = {10.1109/TCOMM.2025.3535864}
}

@inproceedings{11443816,
  author      = {Antonelli, L. and D'Amico, A. A. and Sanguinetti, L.},
  title       = {An enhanced polar-domain dictionary design for elevated {BSs} in near-field {U-MIMO}},
  booktitle   = {Proc. 59th Asilomar Conf. Signals, Syst. Comput.},
  pages       = {1784-1789},
  year        = {2025},
  doi         = {10.1109/IEEECONF67917.2025.11443816}
}

@article{antonelli2026novelnearfielddictionarydesign,
  author      = {Antonelli, Luca and D'Amico, Antonio A. and Sanguinetti, Luca},
  title       = {A novel near-field dictionary design for hybrid {MIMO} with uniform planar arrays},
  journal     = {IEEE Trans. Wireless Commun.},
  year        = {2026},
  note        = {Early access},
  doi         = {10.1109/TWC.2026.3737524}
}

@book{ConwaySloane1999,
  author      = {Conway, J. H. and Sloane, N. J. A.},
  title       = {Sphere Packings, Lattices and Groups},
  series      = {Grundlehren der Mathematischen Wissenschaften},
  edition     = {3rd},
  volume      = {290},
  publisher   = {Springer},
  address     = {New York, NY, USA},
  year        = {1999},
  doi         = {10.1007/978-1-4757-6568-7}
}

@article{Kershner_1939,
  author      = {Richard Kershner},
  title       = {The number of circles covering a set},
  journal     = {Amer. J. Math.},
  volume      = {61},
  number      = {3},
  pages       = {665--671},
  publisher   = {Johns Hopkins University Press},
  year        = {1939},
  url         = {http://www.jstor.org/stable/2371320}
}

\end{document}